\documentclass[%
 reprint,
superscriptaddress,
aps,
prb,
 amsmath,amssymb,
]{revtex4-2}

\setcitestyle{super}

\usepackage{graphicx}
\usepackage{dcolumn}
\usepackage{bm}
\usepackage{booktabs,xcolor,colortbl}
\usepackage{xparse}
\usepackage[toc,page]{appendix}
\usepackage{caption}
\usepackage{subcaption}
\usepackage{float}
\usepackage{afterpage}

\usepackage{tabularx,booktabs}
\newcommand{\br}{{\mathbf{r}}}

\usepackage{hyperref}
\usepackage{todonotes}
\usepackage{color}

\begin{document}

\title{Improved proton-transfer barriers with van der Waals density functionals:
Role of repulsive non-local correlation }

\author{S. Seyedraoufi}
\affiliation{Department of Mechanical Engineering and Technology Management, Norwegian University of
Life Sciences, Norway.}

\author{Kristian Berland}
\email[E-mail: ]{kristian.berland@nmbu.no}
\affiliation{Department of Mechanical Engineering and Technology Management, Norwegian University of
Life Sciences, Norway.}

\date{\today}

\begin{abstract}
Proton-transfer (PT) between organic complexes is a common and important biochemical process. 
Unfortunately, PT energy barriers are difficult to accurately predict 
using density functional theory (DFT); in particular, the generalized gradient approximation (GGA) tends to underestimate PT barriers. 
Moreover, PT typically occurs in environments where dispersion forces contribute to the cohesion of the system; thus, a suitable exchange-correlation functional 
should accurately describe both dispersion forces and PT barriers. 
This paper provides benchmark results for the PT barriers of several density functionals including several variants of the van der Waals density functional (vdW-DF).
The benchmark set comprises small organic molecules with inter- and intra-molecular PT. The results show that replacing GGA correlation with a fully non-local vdW-DF correlation increases the PT barriers, making it closer to the quantum chemical reference values.
In contrast, including non-local correlations with the Vydrov-Voorhis (VV) method or dispersion-corrections at the DFT-D3 or the Tkatchenko-Scheffler (TS) level
has barely any impact on the PT barriers.
Hybrid functionals also increase and improve the energies and the best performance is provided by a hybrid version of the consistent-exchange van der Waals density functional \mbox{vdW-DF-cx}. 
For the formic acid dimer PT system, we analyzed the GGA exchange and non-local correlation contributions. The analysis shows that the repulsive part of the non-local correlation kernel plays a key role in the PT energy barriers predicted with vdW-DF. 
\end{abstract}

\maketitle


\section{\label{sec:level1} Introduction}

Proton transfer (PT) is an ubiquitous chemical reaction and
many biochemical reactions involve PT. 
For instance, proton-coupled charge transfer 
is an established mechanism in enzymology\cite{pt_coupled}
and collective proton transfer in DNA base pairs can give rise to rare tautomers which may lead to mutations.\cite{pt_dna}
In organic solid-state systems, PT can change the nature of the bonding and,
consequently, the properties of a crystal.\cite{bronsted}
PT is also one mechanism for electric polarization switching in the organic ferroelectrics. \cite{pt_ferro_book_chap, Horiuchi1, Horiuchi2} \par

High level quantum chemical methods, in particular, coupled-cluster with single and double and perturbative triple excitations (CCSD(T)) can provide accurate reference data for PT energy barriers, 
but its high computational cost makes it ill-suited for complex PT systems.
The organic complexes in which PT occurs are often held together by van der Waals forces, making it important to assess the accuracy of predicted PT barriers with functionals that include dispersion forces. 

Earlier benchmark studies have found that density functional theory (DFT) in the local density approximation (LDA)\cite{LDA} and generalized-gradient approximation (GGA)\cite{gga_1, gga_2} tend to underestimate PT barriers.\cite{Mangiatordi2012, Barone_proton} Hybrid functionals, which mix in a fraction of Fock exchange,\cite{hybrid, dft_holes, pbe0, b3lyp}
improve performance\cite{Mangiatordi2012} which, in part, can be linked to reduced self-interaction error. \cite{hybrid_sie}
Patchkovskii et al.\cite{Patchkovskii2002} found that Perdew-Zunger self-interaction corrected DFT improves reaction and activation energy barriers for 11 selected ``difficult'' reactions compared to LDA and GGA.
Although self interaction is one source of the PT barrier underestimation, 
lack of non-local correlation effects has also been suggested as a possible source of inaccuracy.\cite{sic}
 \begin{figure}[t!]
     \centering
     \begin{subfigure}[H]{0.11\textwidth}
         \centering
         \includegraphics[width=\textwidth]{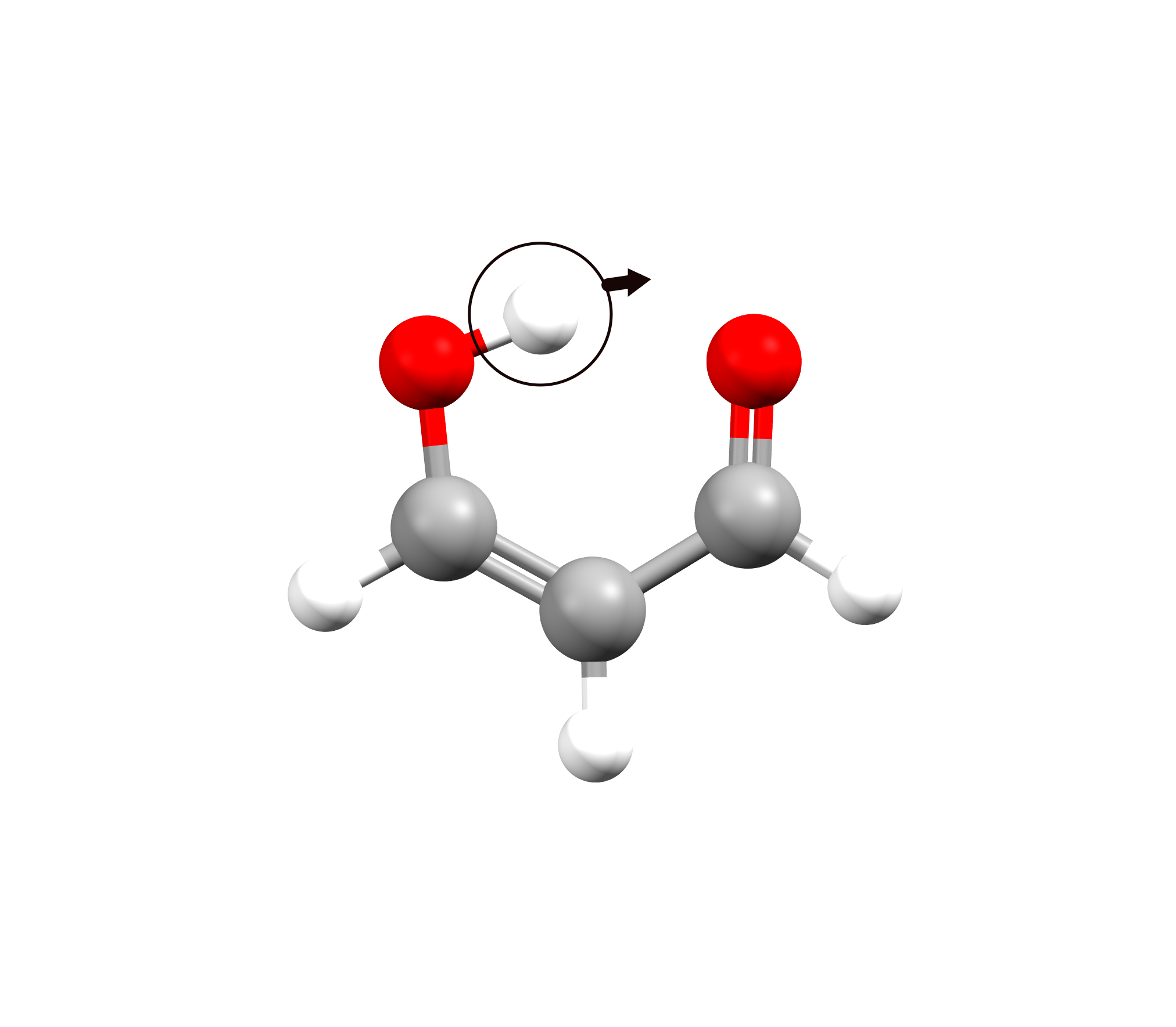}
         \caption{}
         \label{aaa}
     \end{subfigure}
     \begin{subfigure}[H]{0.11\textwidth}
         \centering
         \includegraphics[width=\textwidth]{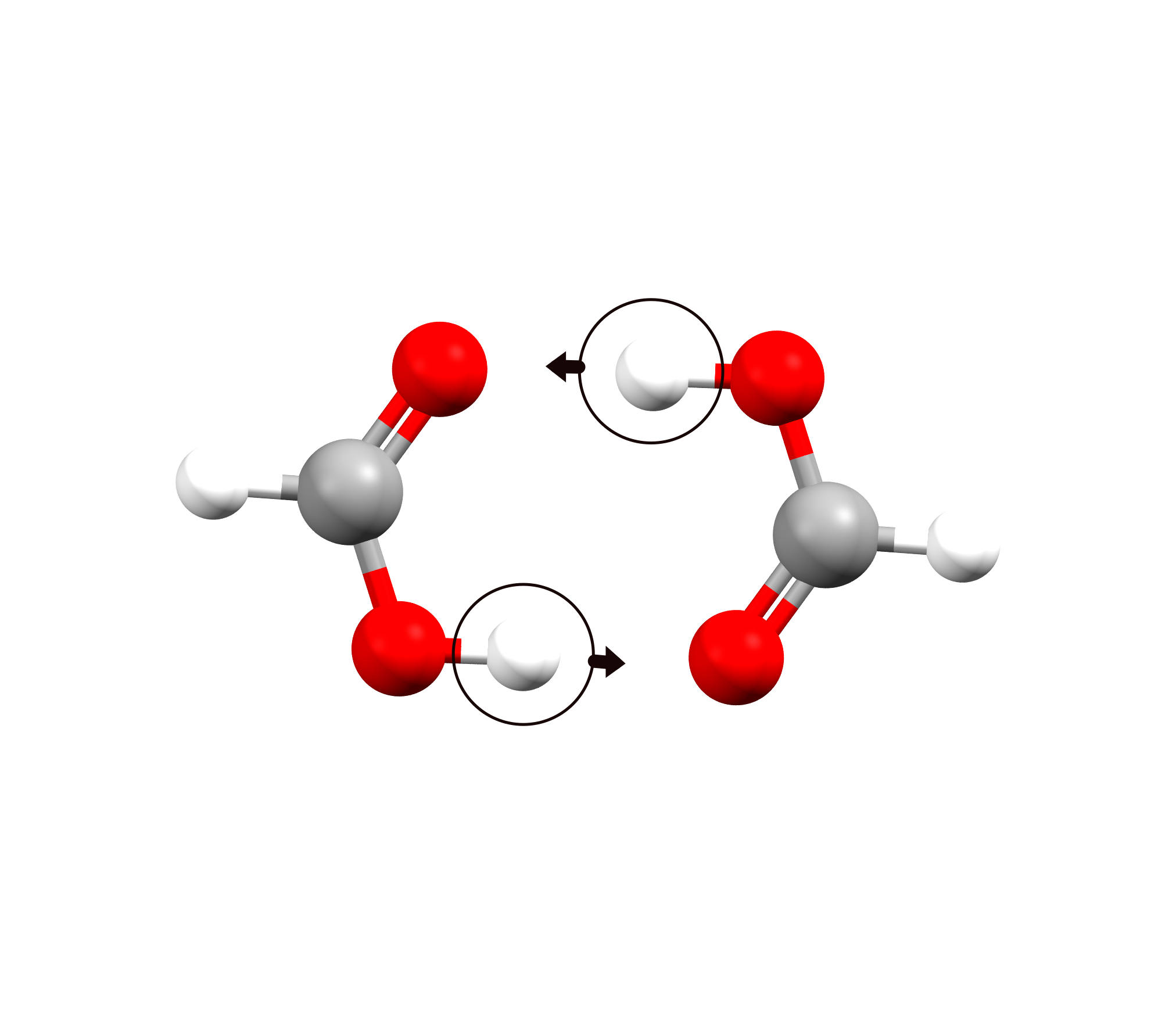}
         \caption{}
         \label{bbb}
     \end{subfigure}
    \begin{subfigure}[H]{0.11\textwidth}
         \centering
         \includegraphics[width=\textwidth]{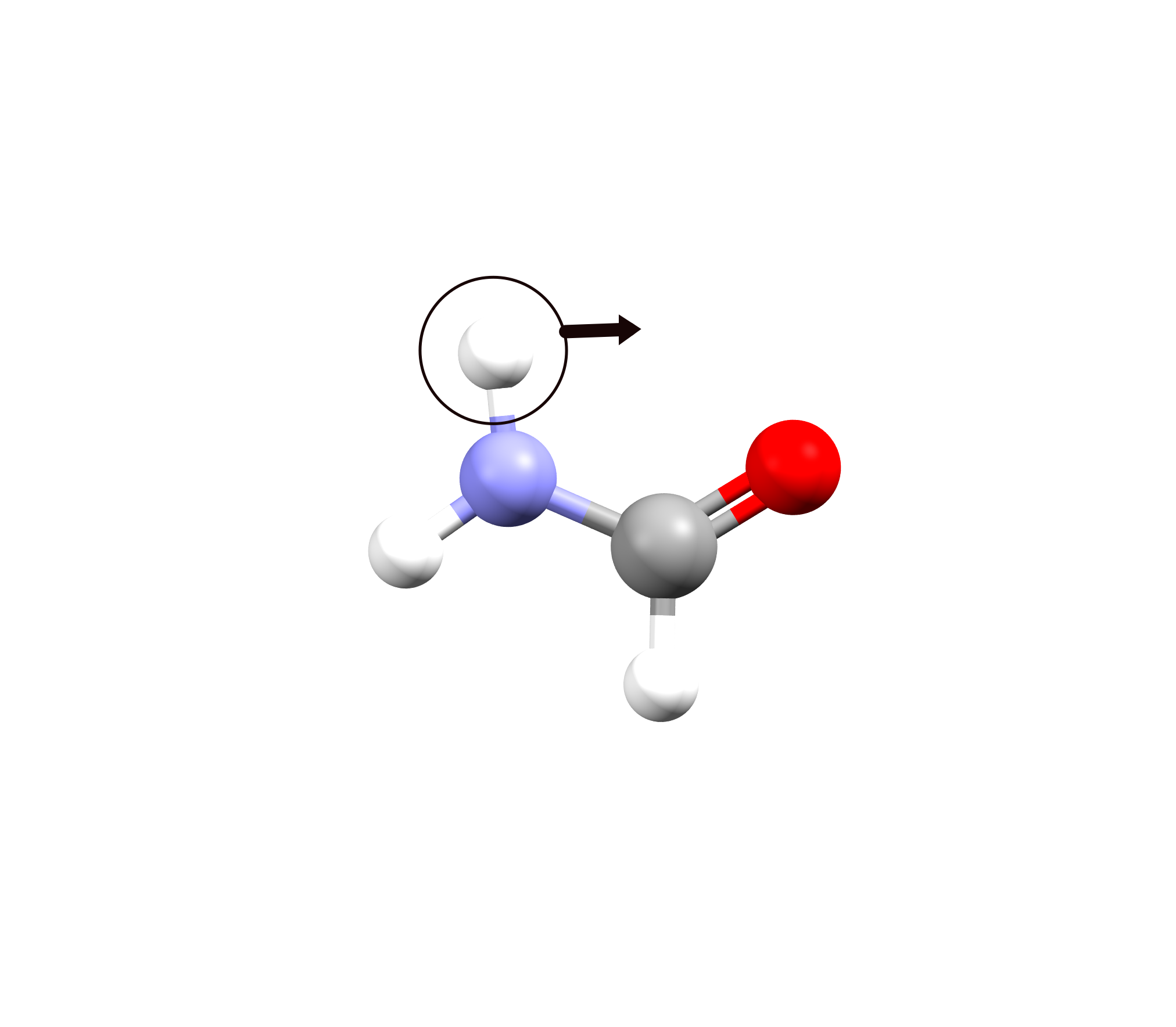}
         \caption{}
         \label{ddd}
     \end{subfigure}
     \begin{subfigure}[H]{0.11\textwidth}
         \centering
         \includegraphics[width=\textwidth]{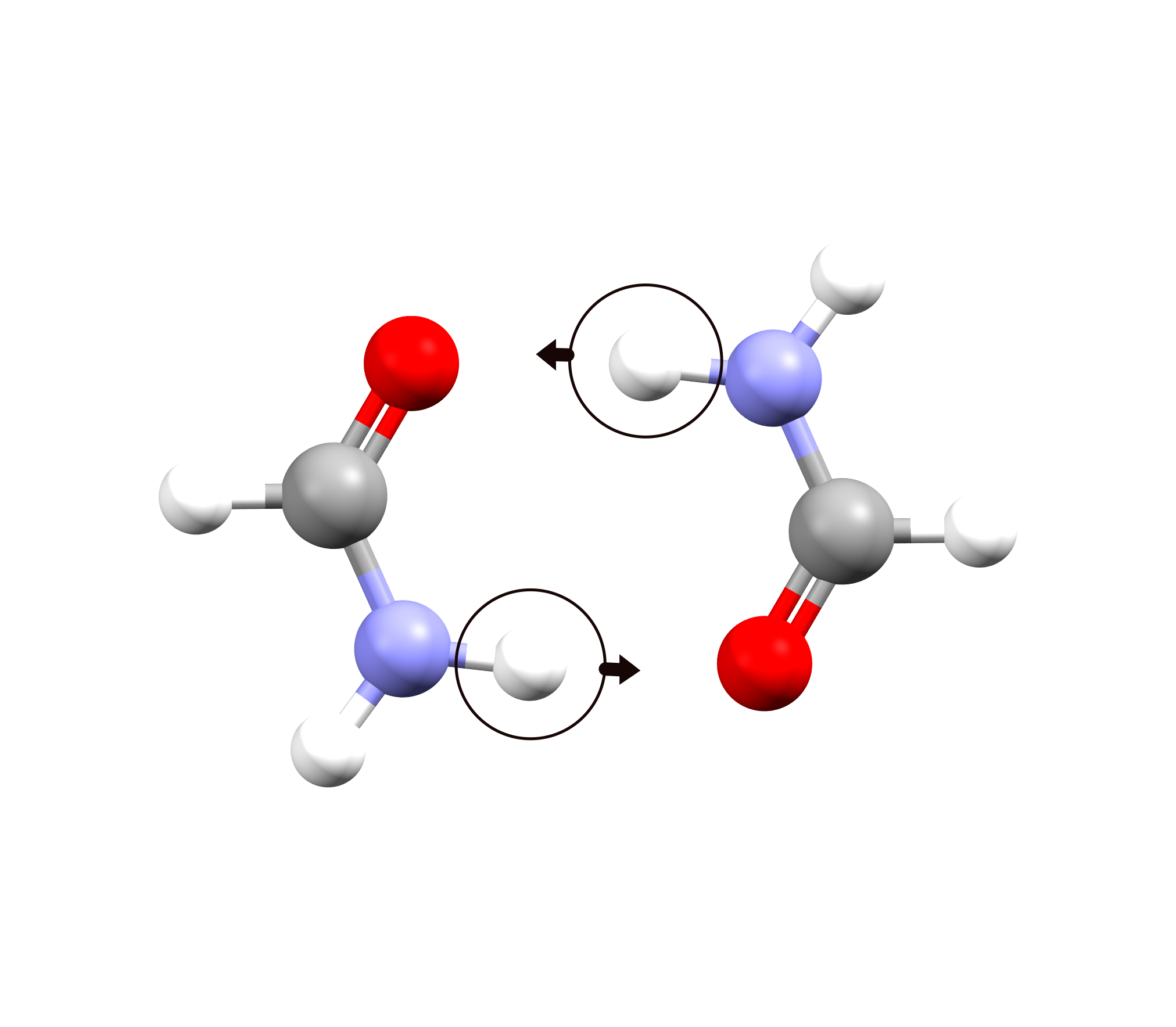}
         \caption{}
         \label{eee}
     \end{subfigure}
          \begin{subfigure}[H]{0.11\textwidth}
         \centering
         \includegraphics[width=\textwidth]{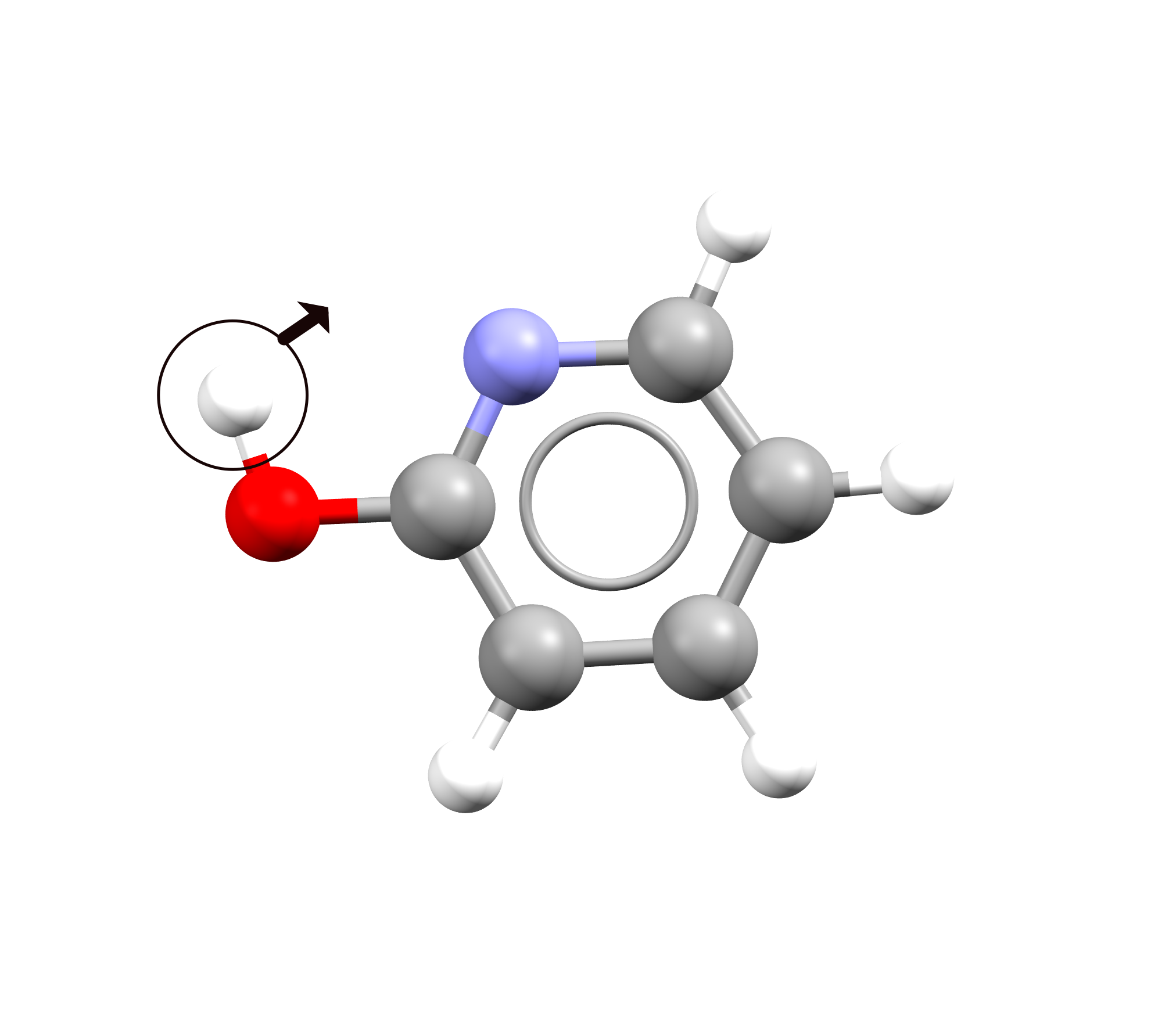}
         \caption{}
         \label{fff}
     \end{subfigure}
     \begin{subfigure}[H]{0.11\textwidth}
         \centering
         \includegraphics[width=\textwidth]{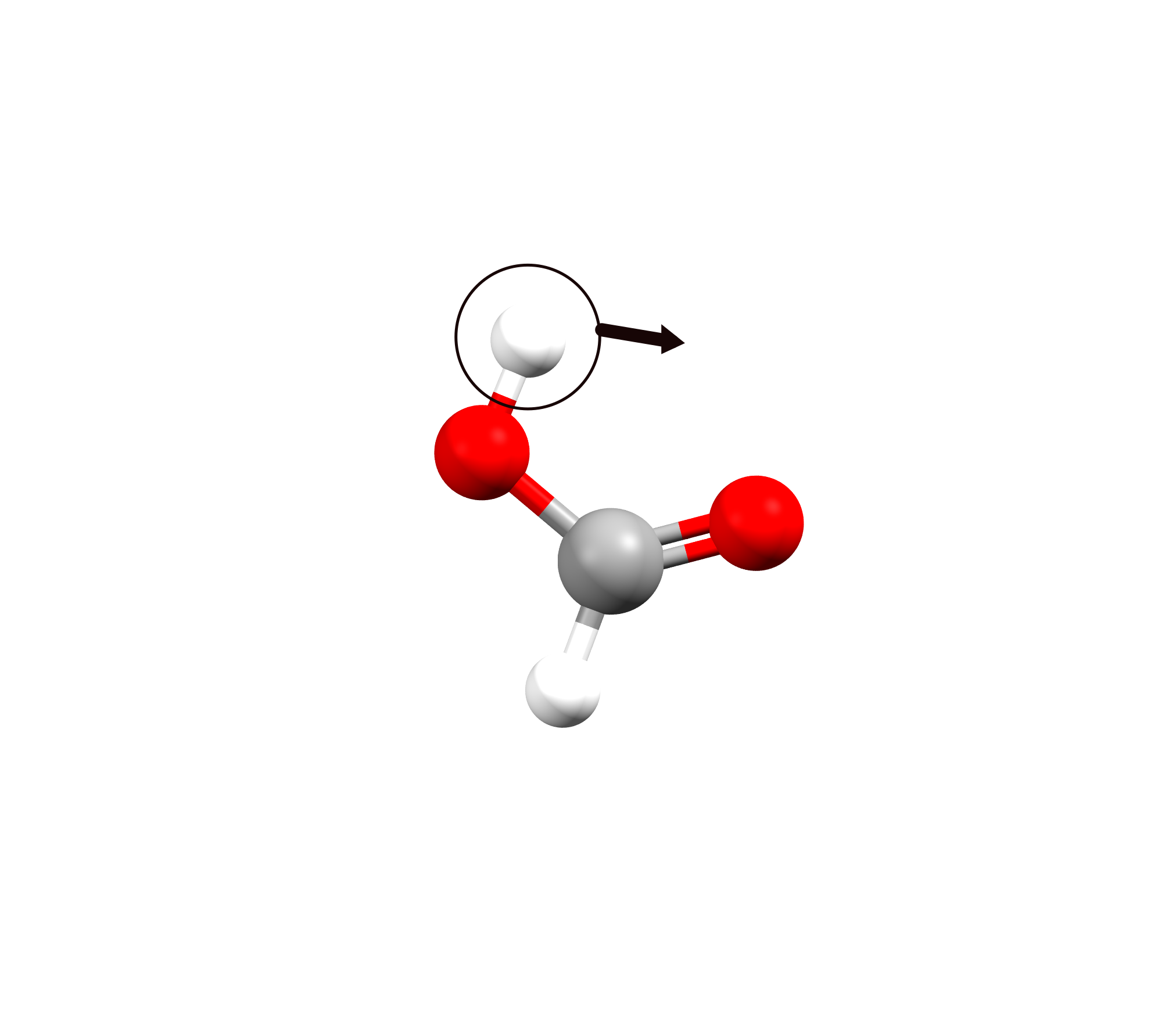}
         \caption{}
         \label{aaa}
     \end{subfigure}
     \begin{subfigure}[H]{0.11\textwidth}
         \centering
         \includegraphics[width=\textwidth]{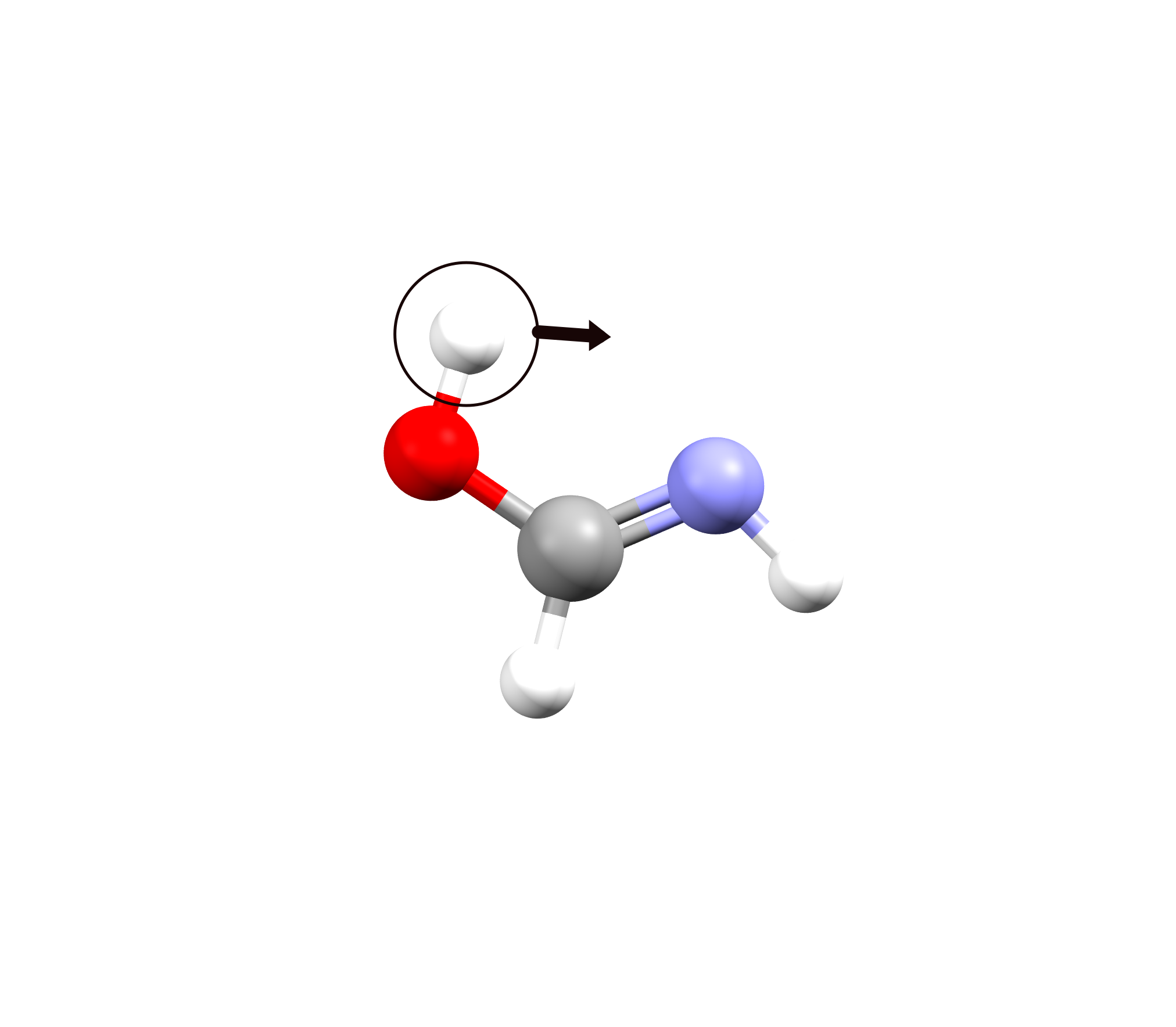}
         \caption{}
         \label{bbb}
     \end{subfigure}
    \begin{subfigure}[H]{0.11\textwidth}
         \centering
         \includegraphics[width=\textwidth]{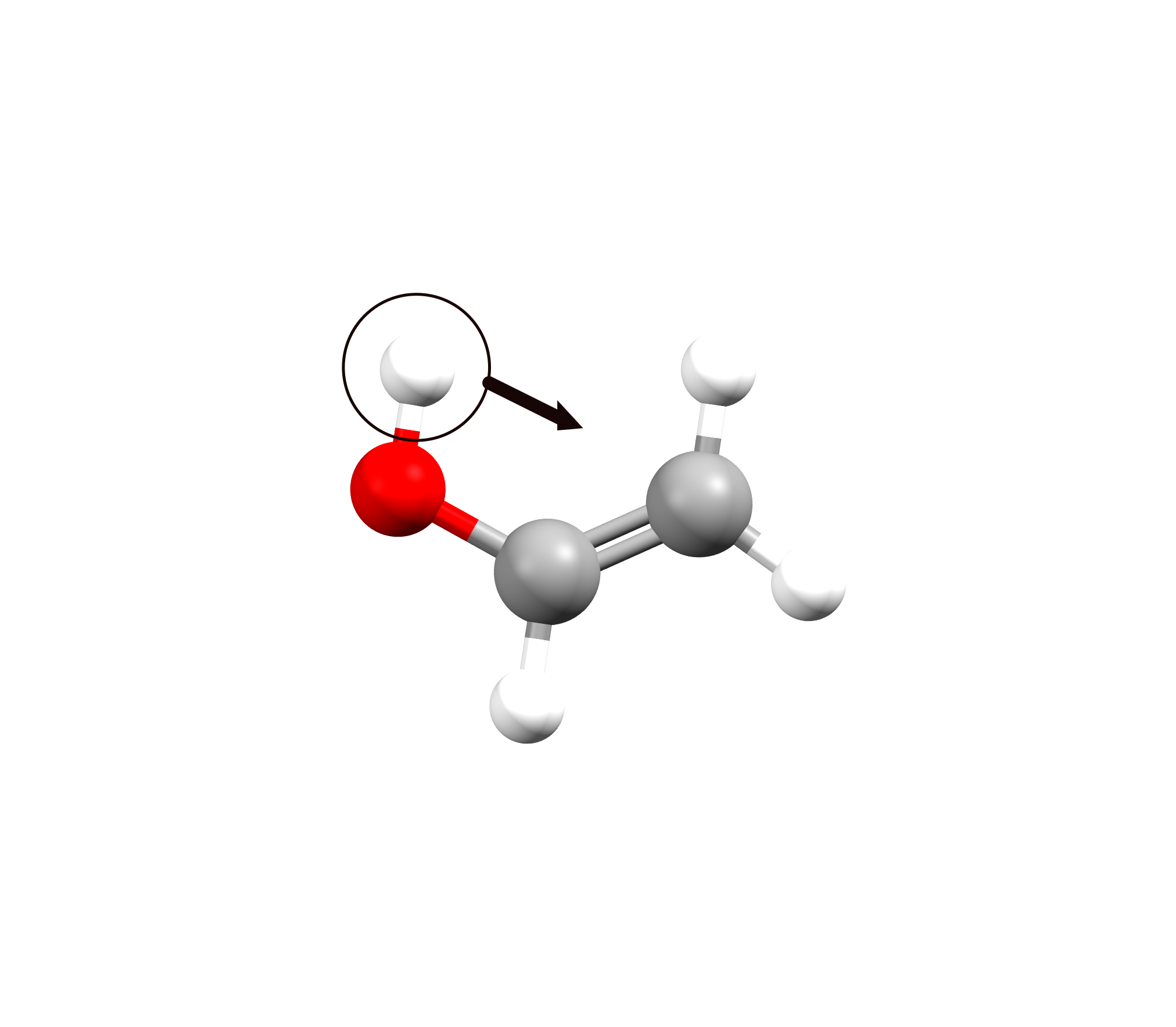}
         \caption{}
         \label{ddd}
     \end{subfigure}
     \begin{subfigure}[H]{0.11\textwidth}
         \centering
         \includegraphics[width=\textwidth]{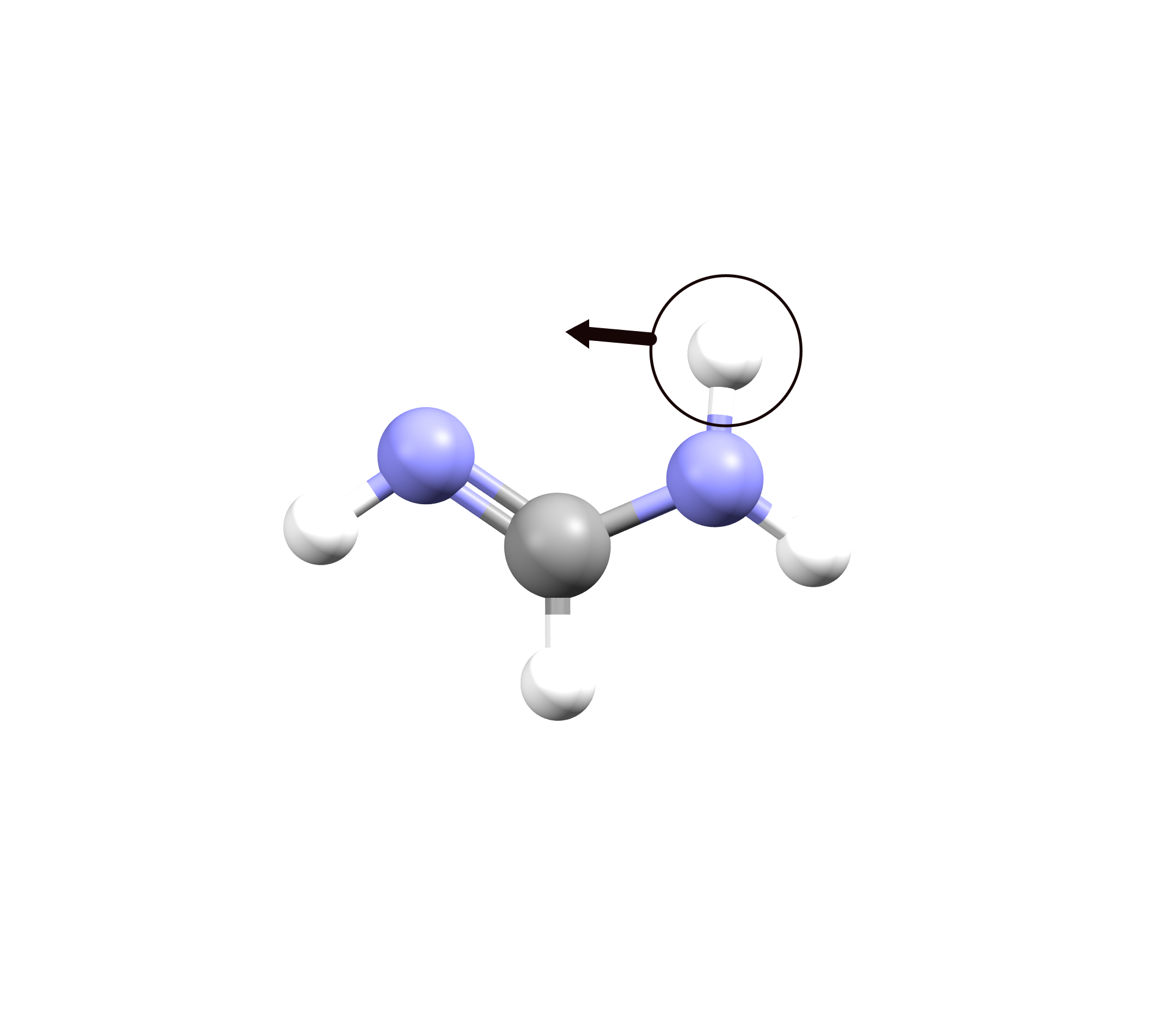}
         \caption{}
         \label{eee}
     \end{subfigure}
          \begin{subfigure}[H]{0.11\textwidth}
         \centering
         \includegraphics[width=\textwidth]{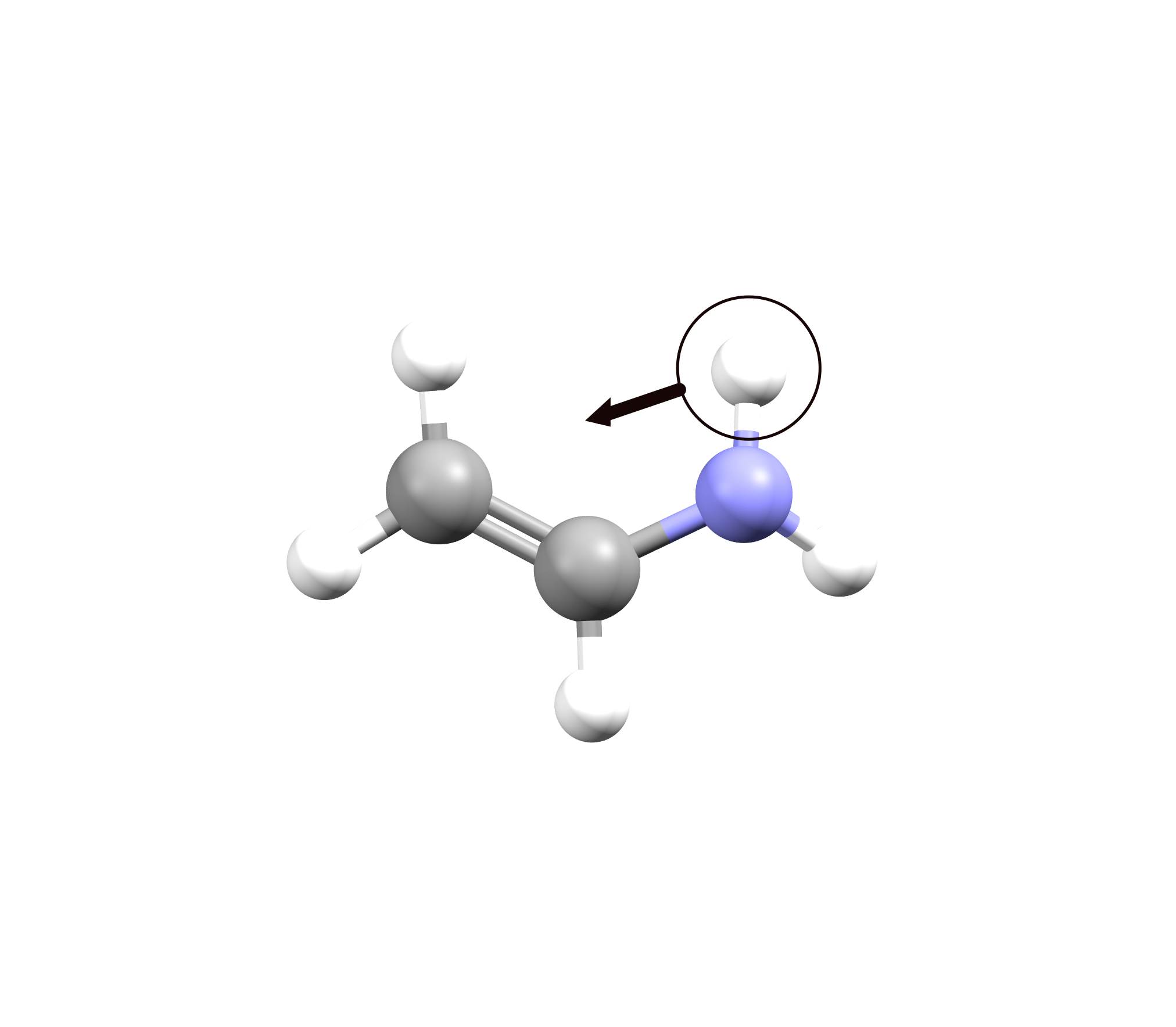}
         \caption{}
         \label{fff}
     \end{subfigure}
          \begin{subfigure}[H]{0.11\textwidth}
         \centering
         \includegraphics[width=\textwidth]{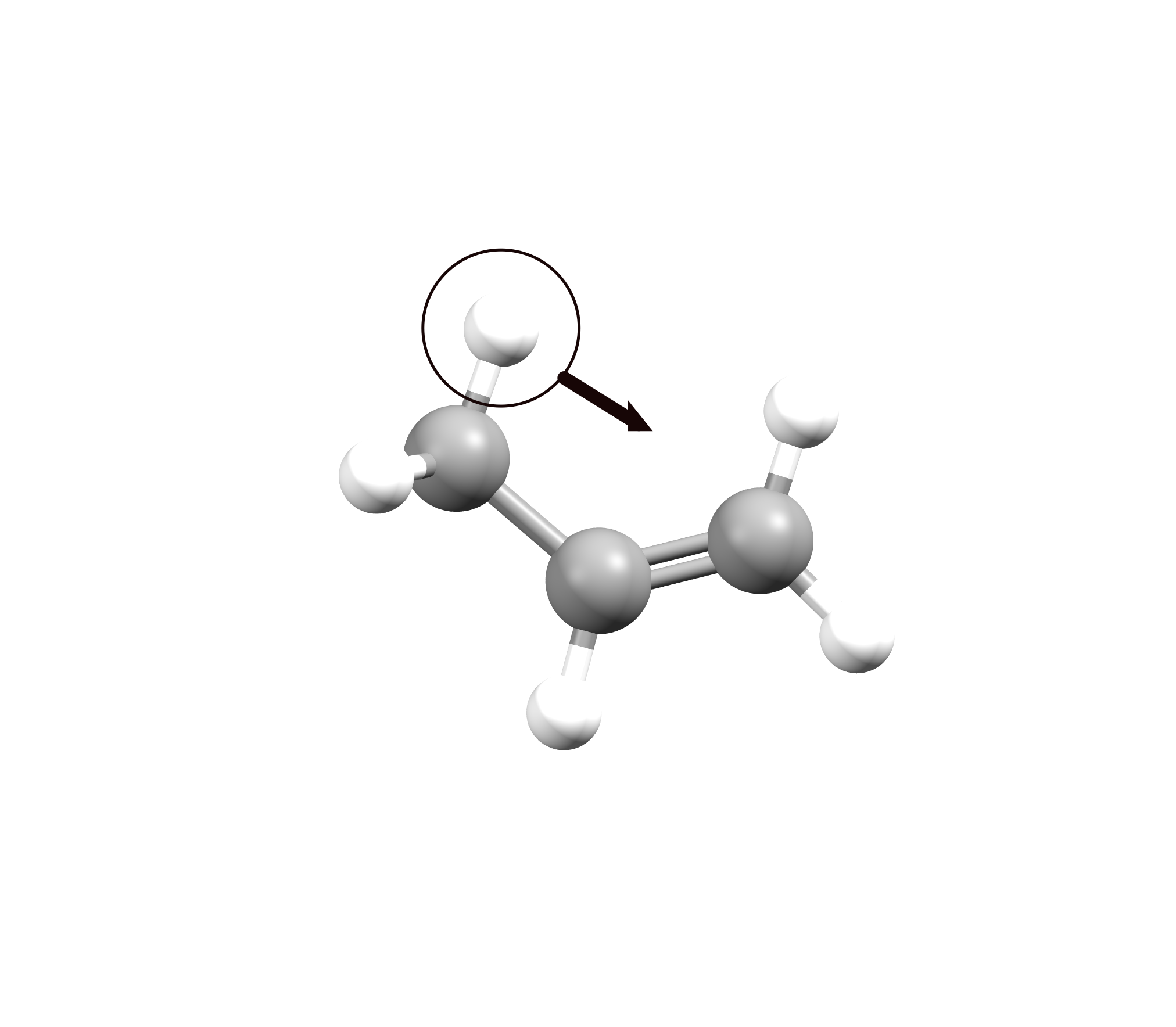}
         \caption{}
         \label{fff}
     \end{subfigure}
          \begin{subfigure}[H]{0.11\textwidth}
         \centering
         \includegraphics[width=\textwidth]{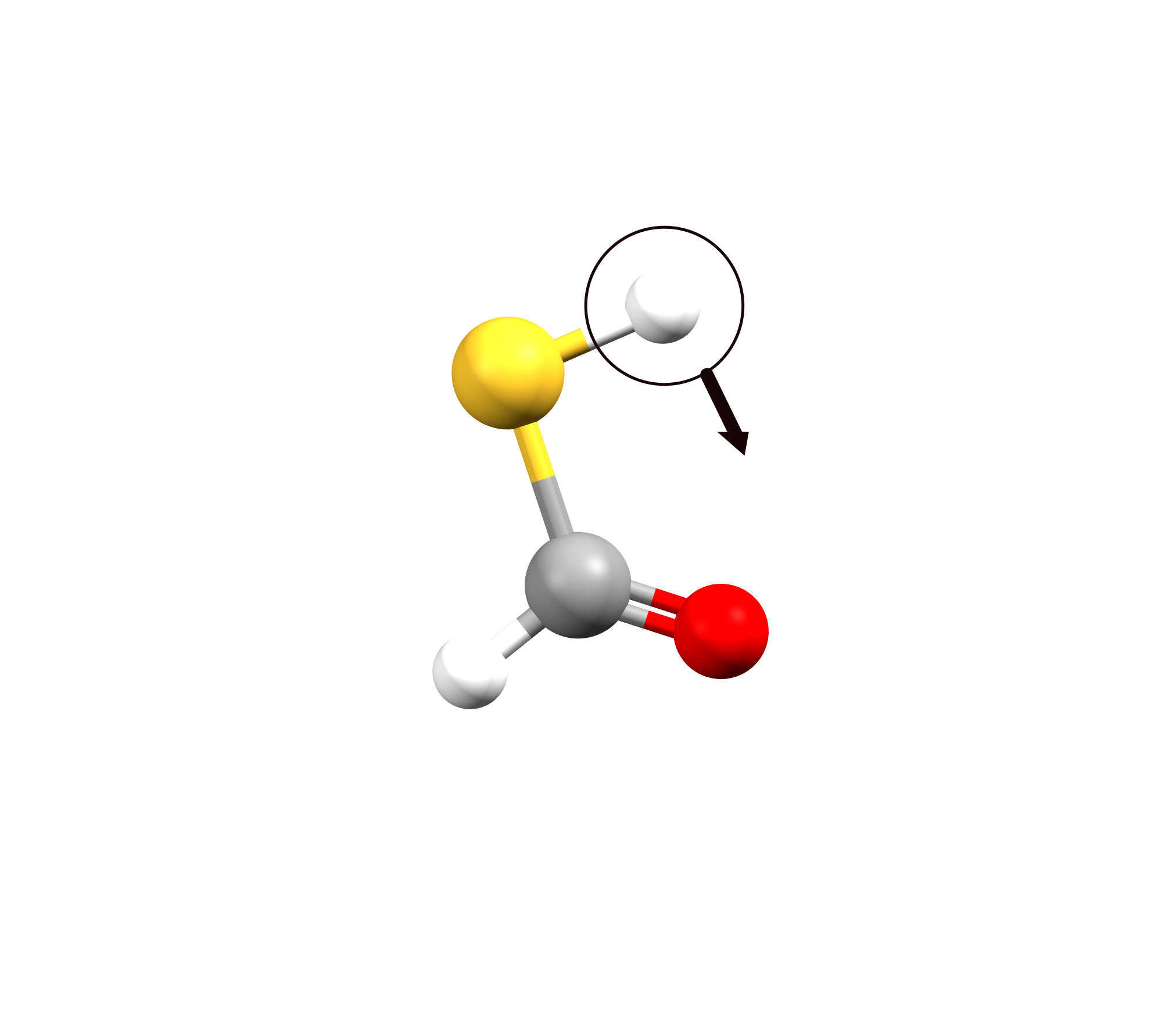}
         \caption{}
         \label{fff}
     \end{subfigure}
          \begin{subfigure}[H]{0.11\textwidth}
         \centering
         \includegraphics[width=\textwidth]{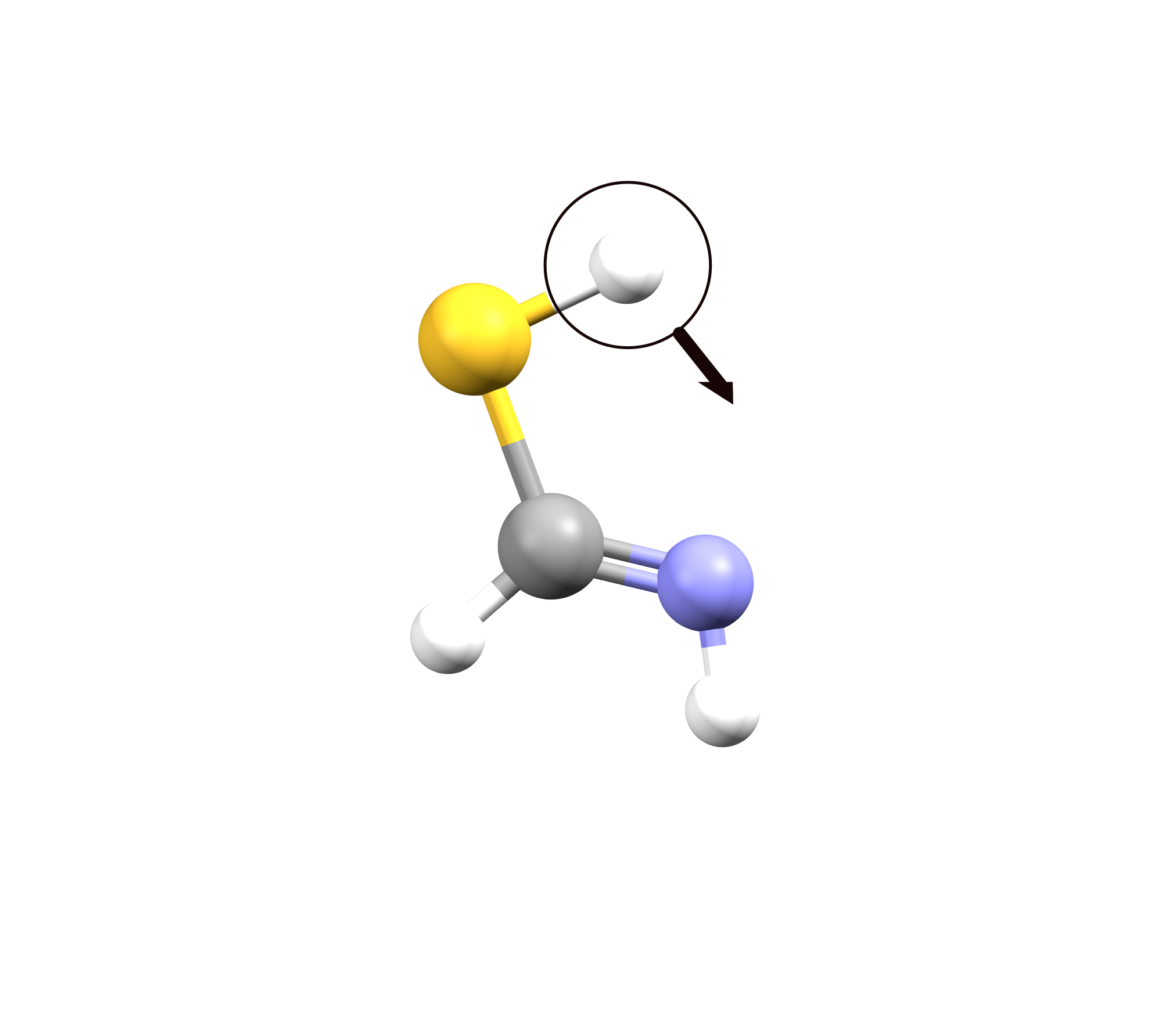}
         \caption{}
         \label{fff}
     \end{subfigure}
          \begin{subfigure}[H]{0.11\textwidth}
         \centering
         \includegraphics[width=\textwidth]{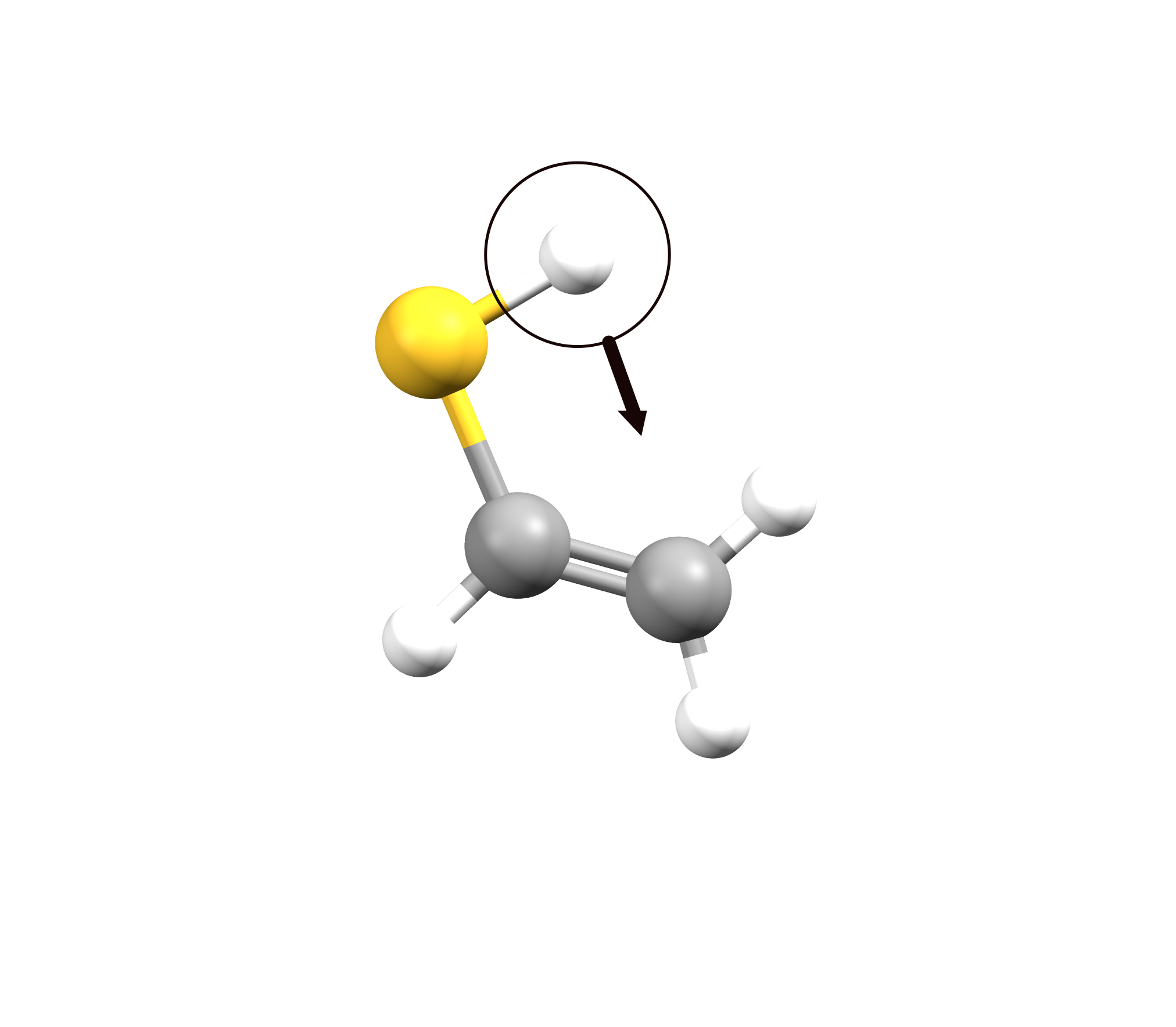}
         \caption{}
         \label{fff}
     \end{subfigure}
        \caption{\raggedright Structures of the PT14 benchmark set. Arrows indicate the direction of proton transfers. Reaction systems comprise: (1) malonaldehyde, (2) formic acid dimer, (3) formamide, (4) formamide dimer, (5) pyridine,  and nine tautomerization reactions including (6-8) carbonyls, (9-10) imines, (11) propene, and (12-14) thiocarbonyls. Color scheme: O (red), C (grey), H (white), N (blue), S (yellow).}
        \label{fig:substructures}
\end{figure}

Several different methods for including dispersion forces in DFT have been developed, but most of these use ``dispersion correction'' scheme in which van der Waals forces are reintroduced with atomistic force-field correction.\cite{ pbe-ts, dft_d, dft_d2, dft_d3, grimme_rev, Sthr2019} 
One class of functionals\cite{vdw, vdw2, revdf2, cx, cx1, opt_b88, optPBE, Chakraborty2020, Thonhauser2007, vdw_book_chap} with a non-local account of the correlation is the Chalmers-Rutgers van der Waals (vdW-DF) density functional\cite{vdw, Thonhauser2007, berland_review}
which is a popular method to describe materials bonded by dispersion forces.
vdW-DF is derived from exact criteria using a plasmon-model of the 
response properties of the electron-gas. \cite{vdw,vdw_book_chap, berland_review, cx1, Hyldgaard2020, interp_vdw}
In the theory, the non-local correlation energy takes the form,
\begin{equation}
E_c^{\rm nl}[n] = \frac{1}{2} \int d^{3}{\mathbf{r}_1} \int d^{3}{\mathbf{r}_2} \, n({\mathbf{r}_1}) \, \phi({\mathbf{r}_1},{\mathbf{r}_2}) \, n({\mathbf{r}_2})\,,
\label{eq:nl}
\end{equation}
in which a kernel function $\phi({\mathbf{r}_1},{\mathbf{r}_2})$ connects two density regions $n({\mathbf{r}_1})$ and $n({\mathbf{r}_2})$. Unlike other dispersion-correction methods, vdW-DF is designed so that $E_c^{\rm nl}$ vanishes seamlessly into a  
homogeneous electron gas limit without the inclusion of damping terms. Therefore, vdW-DF does not include gradient components of the GGA correlation in the total exchange-correlation energy, i.e. 
$E_{xc}[n] = E_x^{\rm{GGA}}[n] +  E_c^{\rm{LDA}}[n] +  E_c^{\rm{nl}}[n] $\,.
Over the years, several variants of vdW-DF \cite{cx1,revdf2, vdw, vdw2, Chakraborty2020, opt_b88, optPBE, vdw_c6} 
have been developed including hybrid variants.\cite{Berland2017,cx0_per} 
vdW-DF has also inspired other non-local correlation functionals; in particular the Vydrov-Voorhis (VV10)\cite{vv10} functional and 
its revision for planewave codes (rVV10).\cite{rvv10} 
While vdW-DF is foremost developed for describing dispersion-bonded systems,
its explicit non-local correlation has also been found to improve various material properties\cite{Hyldgaard2020, revdf2, cx1, erhart1} including image plane states on graphene.\cite{image_pot_hamada} 

This paper provides benchmark results for 22 different functionals including a number of \mbox{vdW-DFs}. 
The benchmark set labelled PT14 is based on the 
9 charge-neutral intra-molecular PT systems by Karton et al.\cite{wcpt}
and 5 inter- and intra-molecular PT systems by Mangiatordi et al.\cite{Mangiatordi2012} 
computed with coupled-cluster method at the CCSD(T) level of theory. The set of systems are displayed in Fig.~\ref{fig:substructures}. 
In addition to \mbox{vdW-DFs}, we tested several GGAs,\cite{pbe, revpbe} two standard
hybrid functionals,\cite{pbe0,b3lyp} and the strongly constrained and appropriately normed (SCAN) meta-GGA functional.\cite{scan}
We also tested the rVV10 variant\cite{vv10, rvv10} as well as the SCAN-rVV10 functional. \cite{scan_rvv10}
Moreover, for several of the GGAs, we tested the effect of adding 
force-field dispersion corrections at the D3\cite{dft_d3} and Tkatchenko-Scheffler (TS)\cite{pbe-ts} 
level of theory. 

We found, as detailed in Sec.~\ref{sec:results}, that \mbox{vdW-DF} tends 
to increase PT barriers compared to GGA and therefore reduce the deviation with the reference data, in particular if the exchange functional is kept fixed. This trend was found to arise from a reduction in the negative correlation contribution to the PT barriers.
To gain better understanding of this result, 
we performed an in-depth analysis of the case of formic acid dimer (system 2 in Fig.~\ref{fig:substructures}), as detailed in Sec.~\ref{sec:analysis}. 
The non-local correlation contribution was analyzed and found to be linked to 
repulsive short-range non-local correlation effects.
This effect is similar to that of GGA-type correlation 
but with the local geometry-sensitivity inherited to \mbox{vdW-DF}. 

\section{\label{sec:method}  Methods}
The benchmark calculations were carried out with the \textsc{VASP} software package,\cite{vasp, vasp1, vasp2} except for the vdW-DF3-opt1, vdW-DF3-opt2,\cite{Chakraborty2020} and B3LYP(-D3)\cite{b3lyp} calculations for which \textsc{Quantum Espresso} \cite{quantum_espresso,Giannozzi_2017} 
was used, as these functionals are only implemented in the latter.  
In the supercells, 15 Å vacuum padding was used to isolate the molecular systems employing the dipole correction scheme of Neugebauer et al.\cite{ldipol} The electronic self-consistency criteria was set to $10^{-6}$~eV. 
The plane-wave energy cutoff was set to $1000$~eV based on our convergence study as detailed in the next section. 
The \textsc{VASP} calculations used hard projected augmented waves (PAW) pseudopotentials (PPs) while the \textsc{Quantum Espresso} used the recently developed optimized norm-conserving Vanderbilt (ONCV).\cite{oncv} 
For the exchange and correlation analysis, we obtained input data from the recently developed \textsc{ppACF} post-processing tool\cite{ppacf} which is distributed as part of the \textsc{Quantum Espresso} software package.

\subsection{Pseudopotential choice and convergence}
\begin{figure}[t!]
\includegraphics[width=\columnwidth]{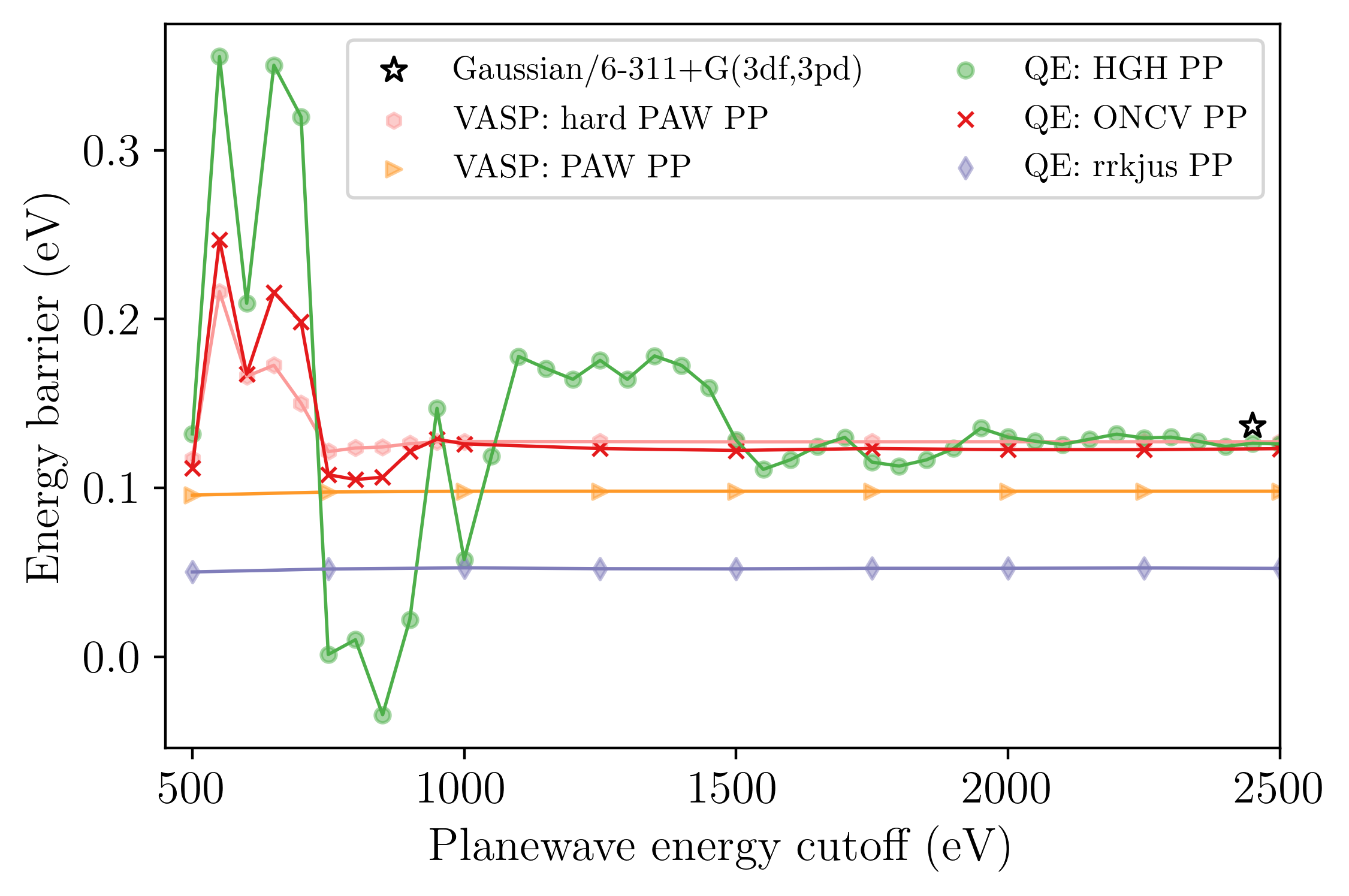}
\caption{\label{fig:pseudop} PT energy barrier sensitivity of the formic acid dimer to planewave energy cutoff for different PPs. The black star indicates the energy barrier of Mangiatordi et al.\cite{Mangiatordi2012} study using a 6-311+G(3df,3pd) orbital basis.}
\end{figure}

In the energy cutoff convergence study and PP selection,
we compared the ultra-soft Rappe-Rabe-Kaxiras-Joannopoulos (RRKJUS) pseudopotential,\cite{rrkj} the hard norm-conserving PP Hartwigsen–Goedecker–Hutter (HGH),\cite{hgh} and the recently developed ONCV as implemented in \textsc{Quantum Espresso} as well as the standard and hard PAW PP\cite{paw, paw_1} as
implemented in \textsc{VASP}. \par
Fig.~\ref{fig:pseudop} displays the results for the case of PT barriers of the formic acid dimer. 
It shows that while ONCV, HGH, and hard PAW converge to a similar value of 0.12~eV, this value differs from the converged value of the ultrasoft RRKJUS and standard PAW, by 
0.07~eV and 0.03~eV, respectively. 
In comparison, the literature value obtained with 6-311+G(3df,3pd) orbital basis set using \textsc{Gaussian} program package\cite{gauss} is 0.13~eV.\cite{Mangiatordi2012} 
The similarity of the ONCV, HGH, and hard PAW results demonstrates the reliability of these PPs. 
Our results are in line with the fact that ONCV has been shown to perform well compared to all-electron results for a selected set of solids including covalent, ionic, and metallic bonding.\cite{oncv} 
In contrast to ONCV and hard PW which converge smoothly 
to within 1~meV at 1000~eV, the PT barriers obtained with HGH fluctuates significantly with the cutoff until the energy differences fall within 1~meV at an energy cutoff of 2800~eV.

\section{\label{sec:results}  Results}

\begin{figure}[h!]
\includegraphics[width=\columnwidth]{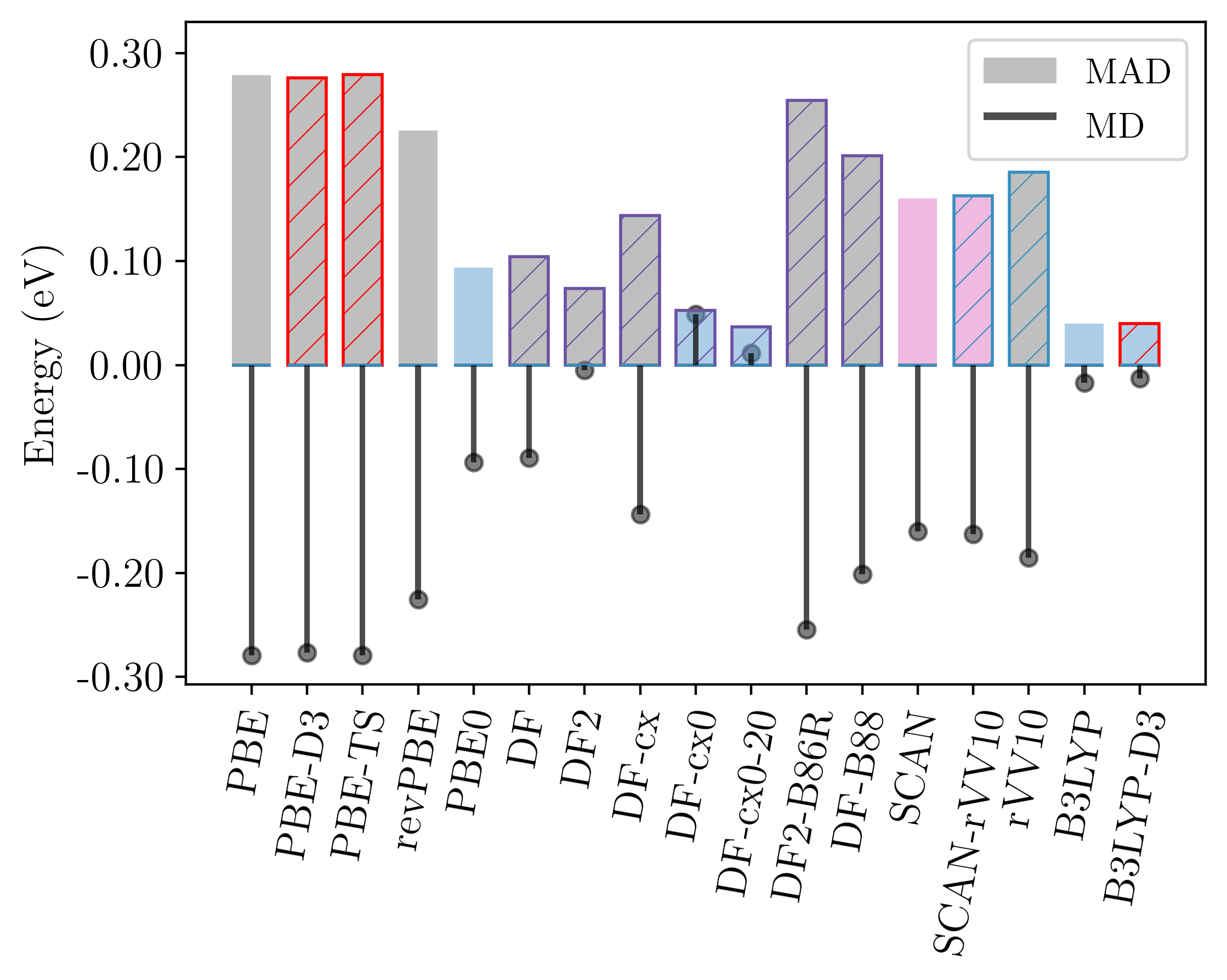}

\caption{\label{fig:functionals} 
Mean (absolute) deviation (M(A)D) for PT barrier energies of the PT14 benchmarking set.
The bar colors indicate the nature of the respective exchange functionals,
with GGAs indicated in gray, hybrids in blue, and meta-GGAs in pink.
Stripes indicate the inclusion of dispersion forces, 
either with a dispersion correction (red stripes), 
or using full non-local correlation within \mbox{vdW-DF} (indigo) or at the rVV10 (light blue) level. (``vdW-'' is dropped in the functional names.}
\end{figure}

Fig.~\ref{fig:functionals} displays computed statistical data for PT barrier energies for the PT14 set for a subset of the benchmarked functionals. 
The full set of results are provided in Tab.~\ref{tab:table_1} of the appendix. Comparing the functionals, we find that the two pure GGAs, revPBE\cite{revpbe} and PBE,\cite{pbe} (indicated by plain grey bars) significantly underestimate the PT barriers. For systems with low PT barriers, underestimating barriers 
can give qualitatively incorrect results. For instance, in the case of the malonaldehyde (system 1 in Fig.~\ref{fig:substructures}), PBE predicts 8~meV far less than the reference value of 168~meV. 
PBE0, which mixes in 25\% Fock exchange with the PBE exchange,\cite{pbe0} improves the PT barriers.
The meta-GGA SCAN is overall more accurate than PBE, but less accurate than PBE0. 
Despite the fact that dispersion forces contribute to 
hydrogen bonding, adding dispersion corrections at the \mbox{DFT-D3}\cite{dft_d3} or at the TS\cite{pbe-ts} level to PBE has almost no impact on the predicted PT barrier. 
Interestingly, a similar insensitivity is also exhibited with the inclusion of rVV10 \mbox{non-local} correlation corrections to SCAN.\cite{scan_rvv10}
In contrast, several \mbox{vdW-DFs}, in particular \mbox{vdW-DF-cx},\cite{cx, cx1} \mbox{vdW-DF},\cite{vdw} 
and \mbox{vdW-DF2}\cite{vdw2} have significantly smaller deviations from the reference than the GGAs. 
The most accurate non-hybrid functional is \mbox{vdW-DF2} with a mean absolute deviation (MAD) of 0.073~eV, only a quarter of that of PBE of 0.279~eV. 
In comparison, the MAD of PBE0 is 0.094~eV. 
The reduced deviation of \mbox{vdW-DF} can be traced to 
the GGA gradient correlations having a larger negative contribution to 
the barrier than the fully non-local \mbox{vdW-DF} correlation. 
For instance, going from revPBE to \mbox{vdW-DF}, which keeps the exchange fixed, 
causes MAD to drop from 0.226~eV to 0.104~eV. 
As both of the non-local correlation and hybrid exchange increase PT barriers, 
the most accurate functional is obtained with the hybrid \mbox{vdW-DF-cx} with 20\% Fock exchange (\mbox{vdW-DF-cx0-20}),\cite{cx0_per, Berland2017} with a MAD of 0.037~eV, while the variant using 25\% Fock exchange (\mbox{vdW-DF-cx0}) gives a MAD of 0.053~eV. 
Finally, we note the empirical B3LYP functional,\cite{b3lyp} which also uses 20\% Fock exchange and a reduced GGA correlation (0.81 LYP \cite{lyp}) also provides accurate PT barriers, with a MAD of 0.041~eV.

\begin{figure}[h!]
\includegraphics[width=\columnwidth]{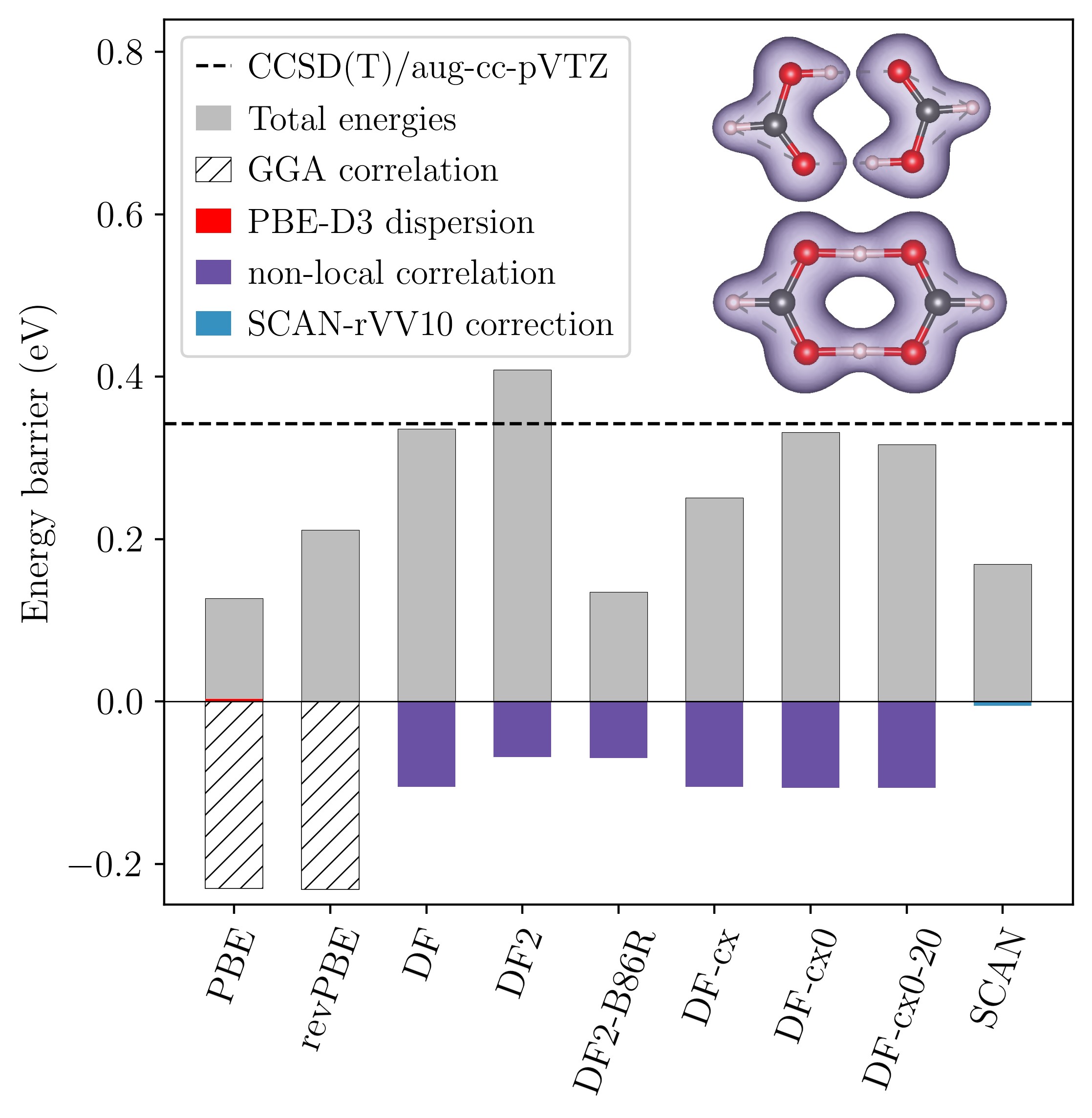}
\caption{\label{fig:formic} Total energies and correlation components of the PT barrier for the formic acid dimer. The quantum chemical reference indicated with the dashed line.\cite{Mangiatordi2012} 
The inset shows the density isosurfaces in the ground and transition state.
The contributions of GGA correlation, non-local correlation with rVV10 and \mbox{vdW-DF}, and D3 corrections are indicated. The visualization in the inset and elsewhere is generated by VESTA.\cite{vesta}}
\end{figure}

Fig.~\ref{fig:formic} shows the individual correlation components 
for eight different functionals 
for the PT barrier of the formic acid dimer. 
It shows how the non-local correlation contribution of \mbox{vdW-DF} reduces the PT barrier
less than the GGA correlation, thus making \mbox{vdW-DF} and \mbox{vdW-DF2} the most accurate 
functionals for this PT barrier. As noted before, the comparison is most evident when comparing 
revPBE and \mbox{vdW-DF} as the exchange is kept fixed. 
The D3 correction in PBE-D3 and the rVV10 non-local correlation contribution in SCAN-rVV10 are barely visible.

\begin{figure}[t!]
\includegraphics[width=\columnwidth]{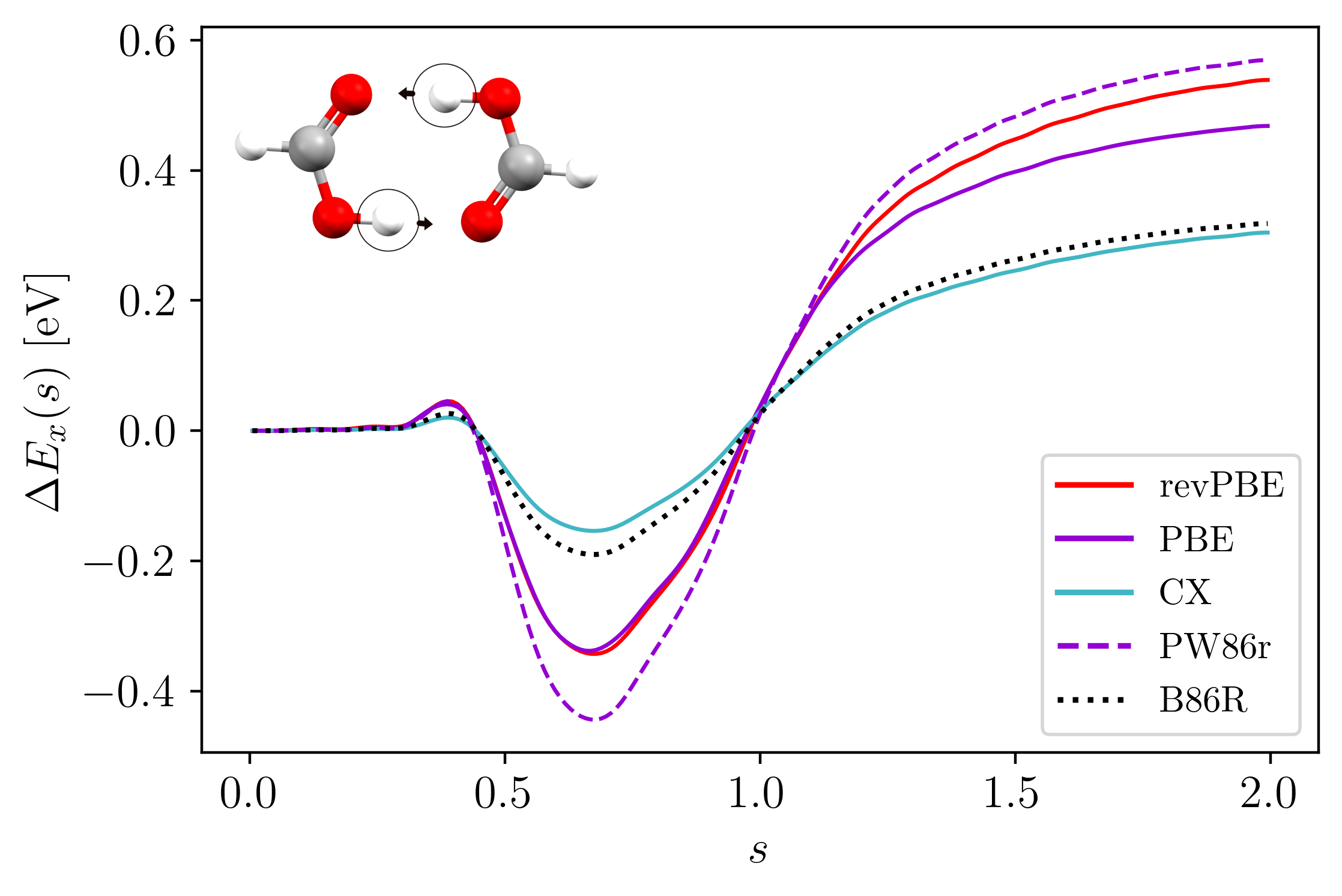}
\includegraphics[width=\columnwidth]{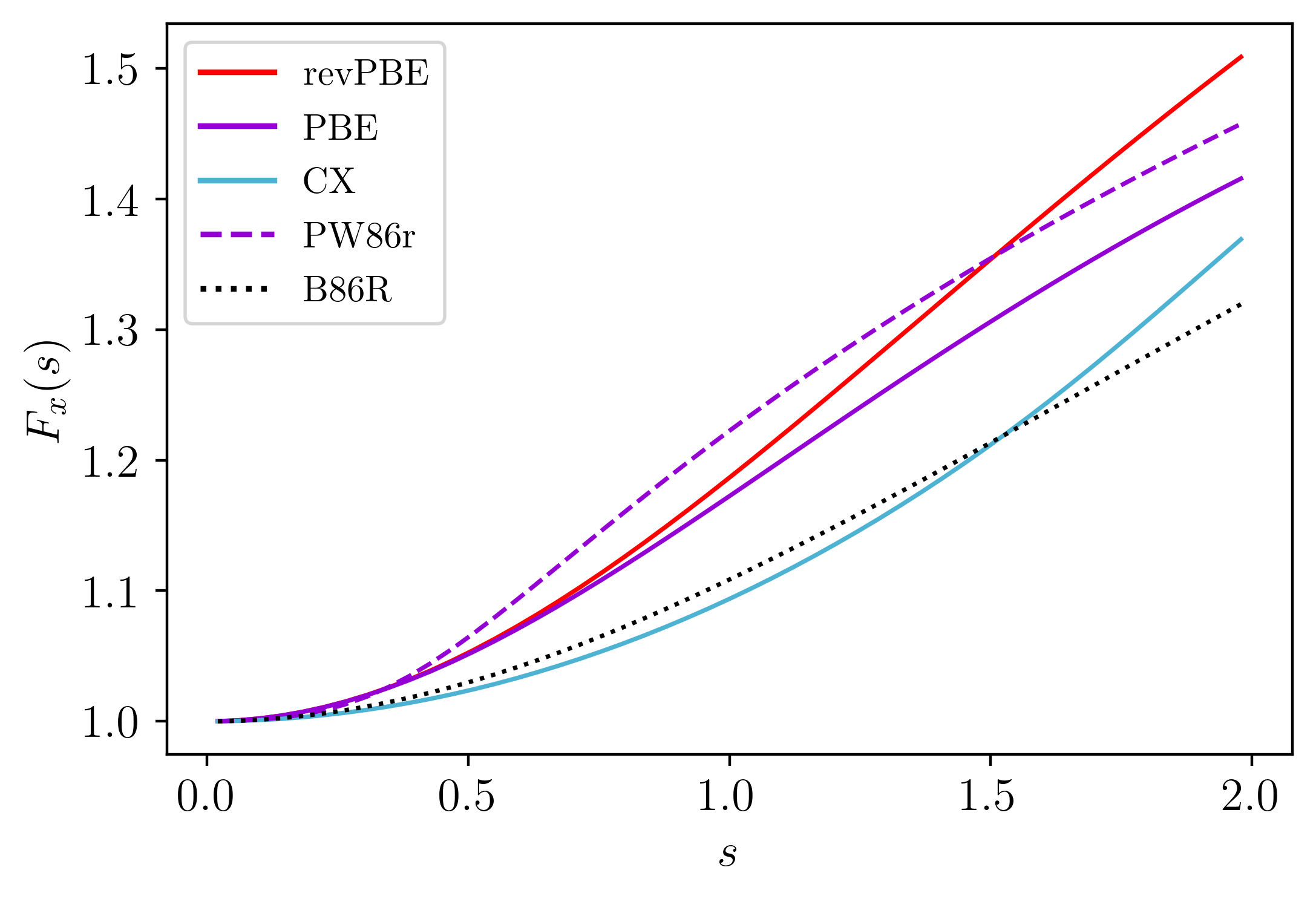}
\caption{\label{fig:delt_e_exchnage} $s$-integrated exchange barrier ($\Delta E_x(s)$) of the formic acid dimer for different selected exchange functionals (upper panel) and Corresponding enhancement factors (lower panel) plotted as a function of $s$.}
\end{figure}

\section{\label{sec:analysis} Analysis of energetic contributions of vdW-DF}

The dramatically improved performance of \mbox{vdW-DF} and \mbox{vdW-DF2} compared to the GGAs, 
and the excellent performance of hybrid \mbox{vdW-DF-cx} prompted us to perform an analysis of the improved 
PT barriers. 
We retain the double-PT formic acid dimer (system 2 in Fig.~\ref{fig:substructures}) as a case study. 

\subsection{Exchange contribution}
The degree of underestimation with GGAs as well as the performance of individual \mbox{vdW-DF} variants depends on the specific exchange functional used.
Jenkins et al.\cite{Jenkins2021} recently highlighted how 
the gradient component of the 
$s$-resolved exchange energy 
can be a useful tool to analyze why the 
choice of exchange enhancement factor 
$F_x(s)$ causes the performance of different \mbox{vdW-DFs} to differ for different types of systems. 
The GGA exchange itself is expressed as modulation of the LDA exchange, as follows
\begin{equation}
E_x^{\rm{GGA}}[n] = \int d^{3}r \, n({\mathbf{r}})  \, \epsilon_x^{\rm{LDA}}(n({\mathbf{r}})) \,F_x(s)\,,
\label{eq:gga_exch}
\end{equation}
with the reduced gradient given by $s({\mathbf{r}}) = |\nabla n({\mathbf{r}})|/2(3\pi^2)^{1/3} n({\mathbf{r}})^{4/3}$. In turn, the $s$-resolved exchange energy is given by 
\begin{align}
	e_x(s) &= \int d^{3}r \, n({\mathbf{r}}) \, \epsilon_x^{\rm{LDA}}(n({\mathbf{r}})) \,[F_x(s) - 1] \,\delta(s - s({\mathbf{r}}))\,.
\label{eq:s_res_ex}
\end{align}
The contribution to the PT energy is given by the difference between the transition state (ts) and ground state (gs) as follows: 
\begin{equation}
\Delta e_x(s) = e_x^{\rm ts}(s) - e_x^{\rm gs}(s)\,.
\label{eq:s_pbe_exc_barrier}
\end{equation}
The $s$-integrated exchange energy density is given by
\begin{equation}
\Delta E_x(s) = \int_0^{s} \Delta e_x(s')  \, ds'\,.
\label{eq:s_pbe_ex_barrier_int}
\end{equation}
Numerically, we computed Eq.~\ref{eq:s_pbe_ex_barrier_int} using the Savitzky–Golay\cite{sgolay} filter fitted with a third-degree polynomial to remove the noise imposed due to grid-point integration
and then take the derivative to obtain Eq.~\ref{eq:s_pbe_exc_barrier}.

The result is shown in Fig.~\ref{fig:delt_e_exchnage}. Note that this analysis only takes the explicit energetic contributions
of Eq.~\ref{eq:s_res_ex} into account, neglecting contributions from self-consistency which can be considerable.\cite{Jenkins2021, Chakraborty2020} 
We can roughly discriminate two main $s$ regions: $0.4 < s < 0.75$ which reduces the PT barrier, and $s>0.75$ which increases the barrier. 
Comparing revPBE to PBE exchange, we can trace the larger PT barrier of the former, to the increasing difference 
between the values of $F_x(s)$ as $s$ increases, and that most of the difference comes from regions with $s>1.0$.
It is also interesting that \mbox{vdW-DF} and \mbox{vdW-DF2} end up with quite similar energies, 
even though the exchange functionals, revPBE and PW86r, differ considerably in shape. 
This result is due to a partial cancellation of positive and negative exchange contributions to the PT barrier. Comparing the enhancement factors, $F_x(s)$, (lower panel Fig.~\ref{fig:delt_e_exchnage}) with $\Delta E_x(s)$ reveals that the larger value of the PW86r $F_x(s)$ coincides with a larger slope of $\Delta E_x(s)$ in most of the negatively and positively-contributing $s$ regions (top panel). 
The final difference also partially cancels with the small difference in non-local correlation contributions. With a MAD of 0.144~eV, 
\mbox{vdW-DF-cx} performs more accurately than \mbox{vdW-DF2-B86R} with a MAD of 0.254~eV. 
This result can be related both to slightly smaller values of $F_x(s)$ in the regime that contributes to lowering the barrier, as well as considerably larger values beyond $s>1.0$.
The full effect of this is partly counteracted by self consistency which influences the kinetic energy considerably in Fig.~\ref{fig:delt_e_exchnage}.
Tab.~\ref{tab:table_1} shows that while several \mbox{vdW-DFs} perform better than GGAs, 
not all \mbox{vdW-DFs} are equally accurate. For instance, PBE performs better than the recently developed vdW-DF3 variants;
however, this is not due to the non-local correlation, but rather their  
``soft'' exchange functionals, i.e. the $F_x(s)$ shape has overall lower values. 
In fact, truly ``soft'' GGAs such as PBEsol \cite{pbesol} are even less accurate 
with a MAD of 0.4~eV. 
The ``soft'' form in the small-to-medium $s$-regime is crucial for making \mbox{vdW-DF} accurate for other classes of systems such as coinage metals and layered systems.\cite{Chakraborty2020, cx1, berland_review, Tran}
In this sense, it is encouraging that
\mbox{vdW-DF-cx}, which can be labeled ``soft'', performs relatively well. Moreover, the most accurate functionals 
are the hybrid variants of \mbox{vdW-DF-cx}, in line with accurate results found for other chemical reactions.\cite{cx0_per, Berland2017}

\subsection{Analysis of the role of non-local correlation}

\begin{figure*}[t!]
\includegraphics[width=0.9\textwidth]{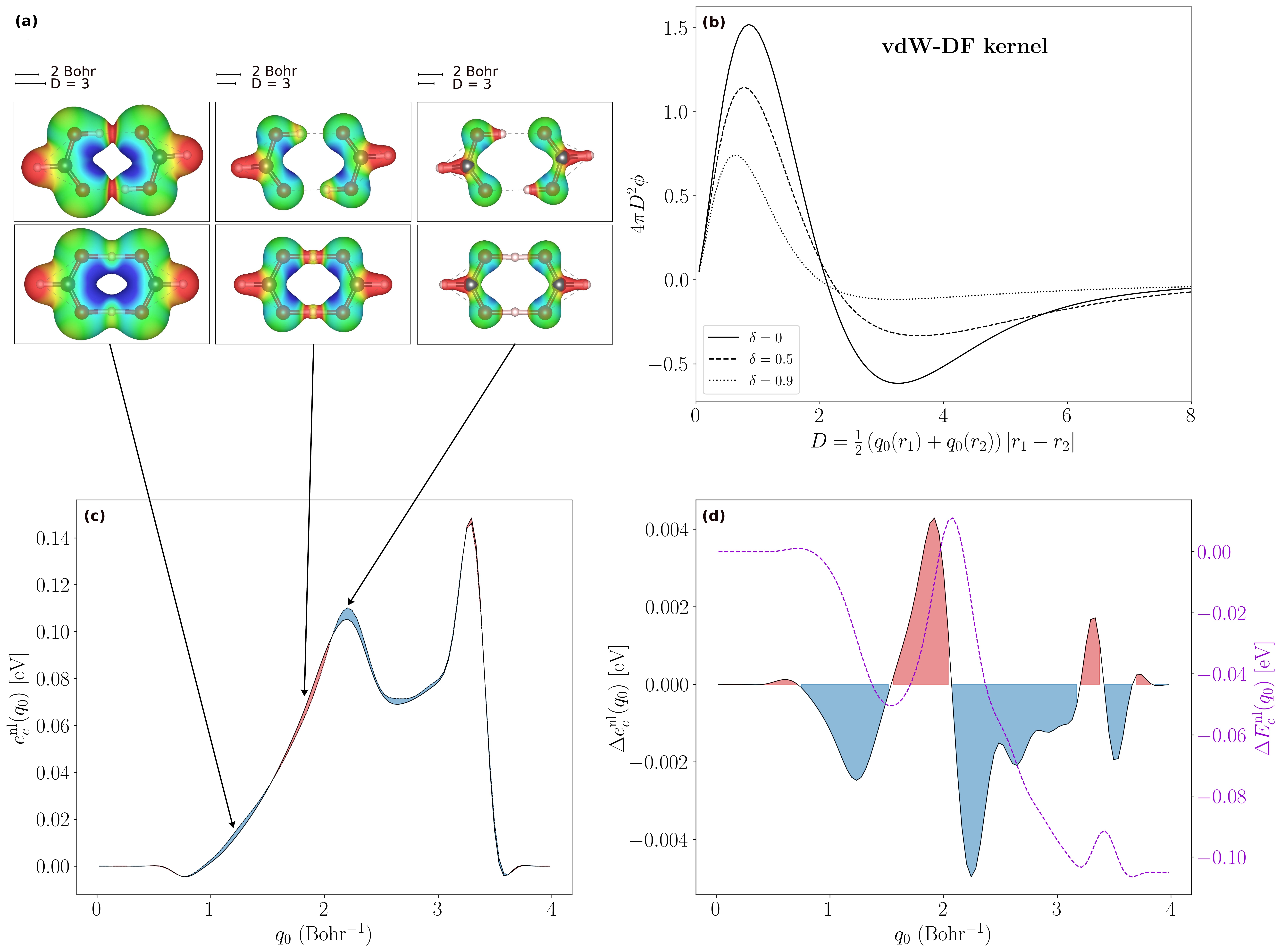}
\caption{\label{fig:kernel} 
Non-local correlation contributions to the double PT transfer energy in formic acid:
({\bf a}) $q_0$-isosurfaces with $q_0 \approx 1.2$, $1.8$, and $2.2$ (atomic units)
	overlaid by non-local correlation density  $e^{\rm nl}_{\rm c}(\br)$ (Eq.~\ref{eq:saptially_nlc}). 
Red indicates negative contributions and blue positive. 
	The upper rulers indicate physical lengths of $2$ Bohr (i.e. $2$ atomic units) $2 \: \rm Bohr \approx 1.06 \: \rm \AA$, and the lower unit-less scaled lengths of $D=3$ entering into the \mbox{vdW-DF} kernel,\cite{vdw} depicted for reference in panel ({\bf b}). 
Panel {(\bf c)} shows the total non-local correlation (Eq.~\ref{eq:q_0_enlc})
of the ground and the transition state. Blue shading indicates negative net contributions to the PT barrier 
and red indicates net positive contributions.
Panel {(\bf d)} shows the corresponding $q_0$-resolved contributions to the PT 
barrier 
	while the dashed curve shows the integrated contributions.} 
\end{figure*}



For analyzing \mbox{vdW-DF}, the local responsivity $q_0({\mathbf{r}})$ is the most natural variable as it enters into the non-local correlation kernel $\phi(\br_1, \br_2) =  \phi(D,\delta )$
through dimensionless parameters $\delta$ and $D$. 
In Sec.~\ref{sec:reduced_pbe_cor}, we will also project non-local correlation density onto $s(\br)$,
for sake of comparison with GGA correlation. 
The parameter $\delta$ is the relative difference in responsivity $q_0(\br)$ (inverse length scale) of two density regions
\begin{equation}
\delta = \frac{|q_0({\mathbf{r}_1}) - q_0({\mathbf{r}_2})|}{q_0({\mathbf{r}_1}) + q_0({\mathbf{r}_2})}\,,
\label{eq:delta}
\end{equation}
and the effective dimensionless separation $D$ is is given by
\begin{equation}
D = \frac{1}{2} \left( q_0({\mathbf{r}_1}) + q_0({\mathbf{r}_2}) \right) |{\mathbf{r}_1} - {\mathbf{r}_2}|\,.
\label{eq:d}
\end{equation}
The parameter $q_0({\mathbf{r}})$
is, within vdW-DF, given by
\begin{equation}
q_0({\mathbf{r}}) = \left( \frac{\epsilon_c^{\rm{LDA}}}{\epsilon_x^{\rm{LDA}}} + 1 - \frac{ \rm Z_{ab}}{9}s({\mathbf{r}})^2 \right) k_{\rm F}({\mathbf{r}})\,,
\label{eq:q_0}
\end{equation}
where $k_{\rm F}({\mathbf{r}})$ is the local Fermi vector. 
${\rm Z_{ab}}$ is equal to $-0.849$ for vdW-DF1 and vdW-DF3-opt1 and $-1.887$ for \mbox{vdW-DF2} and vdW-DF3-opt2. 
The spatial non-local correlation density is given by
\begin{equation}
e_c^{\rm nl}({\mathbf{r}}) = \frac{n({\mathbf{r}})}{2} \int d^3 \mathbf{r}' \, n(\mathbf{r}') \, \phi(\mathbf{r},\mathbf{r}')\,,
\label{eq:saptially_nlc}
\end{equation}
and, subsequently, the $q_0$-resolved non-local correlation can be defined as
\begin{equation}
e_c^{\rm nl}(q_0) = \int d^3 r \, e_c^{\rm nl}({\mathbf{r}}) \, \delta(q_0 - q_0({\mathbf{r}}))\,.
\label{eq:q_0_enlc}
\end{equation}
The $q_0$-resolved non-local correlation contribution to the PT barrier is then given as
\begin{equation}
\Delta e_c^{\rm nl}(q_0) = e_c^{\rm nl, ts}(q_0) - e_c^{\rm nl, gs}(q_0)\,,
\label{eq:q_0_enlc_barr}
\end{equation}
and the corresponding $q_0$-integrated non-local correlation barrier is 
\begin{equation}
\Delta E_c^{\rm nl}(q_0) = \int_0^{q_0} \Delta e_c^{\rm nl}({q_0}') \,d{q_0}'\,.
\label{eq:q_0_integrated_barrier}
\end{equation}

For the analysis of the non-local correlation contributions to PT barriers, Fig.~\ref{fig:kernel} shows four interlinked panels.
Panel {\bf (a)} shows three selected $q_0$ isosurfaces
for the ground state (gs) and transition state (ts).
The overlaid contours indicate
the non-local correlation density as given by Eq.~\ref{eq:saptially_nlc}. 
The three isosurfaces ($q_0 \approx 1.2$, $1.8$, and $2.2$)
were determined from the $q_0$-resolved non-local correlation of Eq.~\ref{eq:q_0_enlc} 
of the ground and transition sates provided in panel {\bf (c)}, with the difference provided in panel ({\bf d}).
The rulers indicate the $D=3$ separation, as well the physical lengths of 2 a.u. (Bohr).
Panel {\bf (b)} shows the \mbox{vdW-DF} kernel. 
The upper left isosurface of panel {\bf (a)} for $q_0 \approx 1.2$
coincides with fairly low density regions, except for the regions directly between 
hydrogens and its oxygen neighbour in the adjacent molecule. Thus, this positive contribution to 
the non-local correlation of the ground state
causes a lowering of the PT barrier. 
The larger blue isosurfaces around the void in between the two molecules
for the transition state
also contribute somewhat to lowering the PT barrier.
This contribution can be viewed as a long-range dispersion effect as the distance from one to the other end of the void coincides with $D\approx3$, i.e. the minimum of the kernel ({\bf b}). 
The isosurfaces corresponding to $q_0 \approx 1.8$ (mid panels {\bf (a)})
show additional lobes around the hydrogen atoms in the transition state, while the ground state isosurface has been disconnected.  
These isosurfaces differences explain the increase in the positive non-local correlation energy of the transition state, thus increasing the magnitude of the PT barrier. 
Finally, the upper right isosurface corresponds to $q_0 = 2.2$ in which the isosurface lobes around the hydrogen in the transition state has vanished, but the isosurfaces around the hydrogen in the ground state cause a significant reduction of the PT barrier.
At this value of $q_0$, the effective separation $D$ has contracted significantly causing intra-molecular 
correlation to be dominating contribution. 

\subsection{\label{sec:reduced_pbe_cor}Reduced-gradient resolved correlation comparison}

\begin{figure}[h]
\includegraphics[width=\columnwidth]{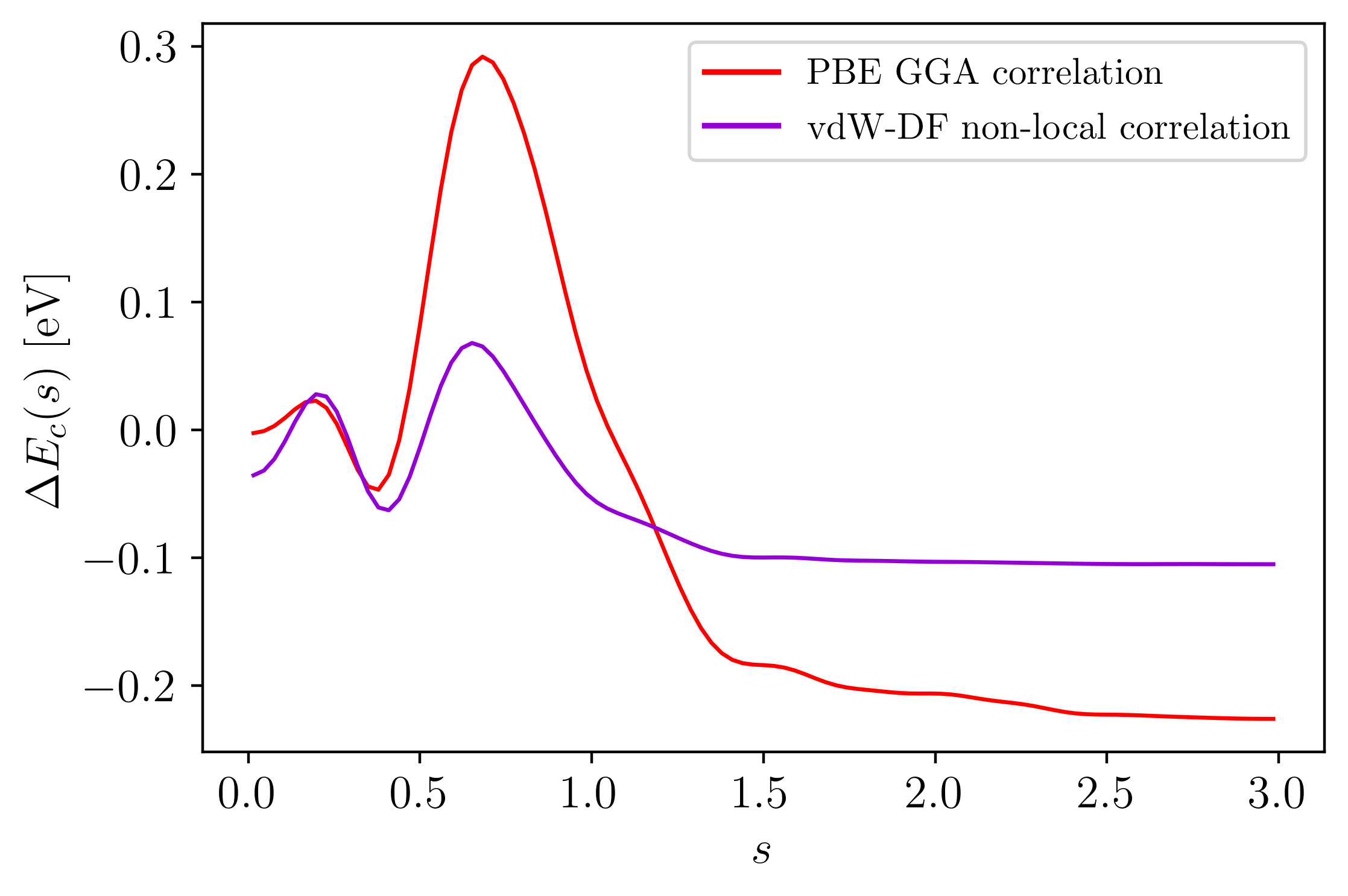}
\caption{\label{fig:pbe_corr} $s$-resolved gradient component of the PBE correlation PT barrier and \mbox{vdW-DF} non-local correlation are plotted as function of $s$.}
\end{figure}

The previous subsection highlighted the utility of 
analyzing \mbox{vdW-DF} in terms of the spatial distribution of $q_0(\br)$ which enters into the kernel $\phi(D, \delta)$ of \mbox{vdW-DF}.
In GGA correlation,\cite{pbe} a key variable is the reduced gradient
$t = |\nabla n({\mathbf{r}})|/2k_{\rm s}({\mathbf{r}})n({\mathbf{r}})$,
which is defined using Thomas-Fermi screening length
$k_{\rm s} = \sqrt{4k_{\rm F} / \pi}$.  
In terms of $t$, the total PBE correlation energy can be expressed
\begin{align}
E_c^{\rm PBE} & = \int d^3 r \, n({\mathbf{r}}) \, [\epsilon_c^{\rm{LDA}}({\mathbf{r}}) + H(n(\br), t(\br))]\,,
\label{eq:pbe_correlation}
\end{align}
where $H$ is the gradient contribution function.
In order to analyze the effect of correlation and exchange on an equal footing,
whether at the GGA or \mbox{vdW-DF} level, we benefit from resolving these quantities onto 
the same variable. We here choose the reduced gradient $s$ which was used for exchange in Fig.~\ref{fig:delt_e_exchnage}.  
The corresponding $s$-resolved differential PBE correlation energy is given by
\begin{equation}
e_c^{\rm PBE}(s) = \int d^3 r \, n({\mathbf{r}}) \, H(r, t(s)) \, \delta(s - s({\mathbf{r}}))\,,
\label{eq:s_pbe_correlation}
\end{equation}
with the corresponding integrated quantity given by
\begin{equation}
\Delta E_c^{\rm PBE}(s) = \int_0^{s} \Delta e_c^{\rm PBE}(s') \, ds'\,.
\label{eq:s_pbe_correlation_barrier_int}
\end{equation}
The projection of \mbox{vdW-DF} onto $s$ is similar to Eq.~\ref{eq:q_0_enlc}. Fig.~\ref{fig:pbe_corr} shows the result for the PT energy barrier contribution 
$\Delta E_c^{\rm PBE}(s)$ and the non-local correlation.
Comparing the curves with the exchange part of Fig.~\ref{fig:delt_e_exchnage}
shows that the curve bear some resemblance, but with opposite prefactor and earlier onset of correlation contribution than exchange with increasing $s$. 
Comparing GGA and \mbox{vdW-DF} correlation
shows that while the shapes are similar, the former has much larger positive and negative 
contributions beyond $s\approx 0.4$. Moreover 
beyond $s\approx 1.5$, \mbox{vdW-DF} correlation flattens, while the magnitude of GGA correlation continues to grow. Inspecting the isosurfaces, we found that the large $s>1$ values corresponds to large isosurfaces similar to the low $q_0$ values.


\section{\label{sec:conclusion}Conclusions and outlook}
We have compared and analyzed the performance of different 
functionals for proton-transfer energy barriers 
and found that using non-local correlation in \mbox{vdW-DF} rather than GGA correlation
 causes a non-intuitive lowering of the energy barriers, typically improving accuracy compared to GGA, which tends to underestimate energy barriers. 
The best performance was provided by a hybrid version of the consistent-exchange van der Waals density functional \mbox{vdW-DF-cx}
with 20\% Fock exchange. 

The improved PT transfer barriers of \mbox{vdW-DF} is highly encouraging
because these functional can account for dispersion forces
responsible for the cohesion of many proton-transfer systems.
Beyond this, our study points to the possibility that the usage of a non-local kernel 
reflects a true geometry-sensitive repulsive short-range correlation contribution. 
In this context, we emphasize that while \mbox{vdW-DF} was designed with dispersion forces in mind,
the theory is rooted in exact constraints and many-body theory and therefore 
is not limited to long range dispersion forces as such. \mbox{vdW-DF} is a true non-local correlation functional with contributions 
both at short and long ranges. 
At the same time, the performance of approximative exchange-correlation functionals 
are generally contingent on the various simplifications and parametrizations used in their construction. The improved performance may also be a mostly ``fortuitous'' effect arising due to typically smaller semi-local correlation-type effects in \mbox{vdW-DF}.
Exploring this question merits further 
theoretical and computational investigations and may pave the way for more accurate methods 
using non-local functionals for improving chemical reaction energies. 


\begin{acknowledgments}
The computations of this work were carried out on UNINETT Sigma2 high performance computing resources (grant NN9650K). This work is supported by the Research Council of Norway as a part of the Young Research Talent project FOX (302362). We thank Rasmus Tranås and Elin Dypvik Sødahl for comments.  
KB acknowledges discussions with T. Jenkins and T. Thonhauser on visualization.
\end{acknowledgments}

\appendix

\section{Tabulated data for PT14 set}
\begin{table*}[]
\caption{\label{tab:table_1} PT energy barriers, mean absolute deviations (MAD), mean deviations (MD), mean absolute relative deviation (MARD), and mean relative deviations (MRD) for the PT14 benchmarking set for different density functionals. The units are in electronvolts (eV). }
\begin{ruledtabular}
\begin{tabular}{*{13}{p{1.3cm}}}
Systems & Ref. &  PBE &  PBE-D3 &  PBE-TS &  revPBE &  PBE0 &  SCAN &  SCAN-rVV10 &  rVV10 &  B3LYP &  B3LYP-D3 & PBEsol \\
\hline
\midrule
1  & 0.168 & 0.008 &   0.015 &   0.009 &    0.049 & 0.073 & 0.038 &       0.036 &  0.073 &  0.130 & 0.144 & -0.079\\
2 &  0.343 & 0.127 &   0.130 &   0.130 &    0.211 & 0.207 & 0.160 &       0.155 &  0.208 &  0.289 &         0.289 & -0.029 \\
3  &2.028& 1.814 &   1.818 &   1.813 &    1.865 & 1.999 & 1.950 &       1.948 &  1.925 &  2.080 &         2.087 & 1.692 \\
4  & 0.828& 0.655 &   0.657 &   0.650 &    0.762 & 0.755 & 0.707 &       0.699 &  0.780 &  0.885 &         0.886 & 0.451\\
5  & 1.650& 1.328 &   1.326 &   1.324 &    1.369 & 1.540 & 1.472 &       1.471 &  1.406 &  1.589 &         1.587 &  1.228\\
6  &1.594& 1.347 &   1.352 &   1.348 &    1.391 & 1.536 & 1.446 &       1.445 &  1.434 &  1.597 &         1.606 & 1.242\\
7  &1.570 & 1.271 &   1.272 &   1.271 &    1.319 & 1.467 & 1.408 &       1.406 &  1.348 &  1.520 &         1.521 & 1.163\\
8  &2.643& 2.254 &   2.255 &   2.254 &    2.287 & 2.495 & 2.408 &       2.406 &  2.351 &  2.559 &         2.562 & 2.152 \\
9  &2.061& 1.778 &   1.778 &   1.776 &    1.832 & 1.973 & 1.925 &       1.923 &  1.881 &  2.046 &         2.045 & 1.654 \\
10 &2.848& 2.509 &   2.510 &   2.508 &    2.544 & 2.741 & 2.673 &       2.670 &  2.625 &  2.818 &         2.820 & 2.403\\
11 &3.523& 3.144 &   3.144 &   3.147 &    3.174 & 3.472 & 3.370 &       3.368 &  3.224 &  3.524 &         3.523 & 3.060\\
12 &1.388& 1.171 &   1.178 &   1.172 &    1.237 & 1.338 & 1.265 &       1.262 &  1.270 &  1.435 &         1.448 & 1.037\\
13 &1.256& 0.987 &   0.991 &   0.986 &    1.052 & 1.158 & 1.089 &       1.087 &  1.078 &  1.248 &         1.254 & 0.854 \\
14 & 2.55& 2.150 &   2.156 &   2.152 &    2.200 & 2.387 & 2.298 &       2.295 &  2.251 &  2.487 &  2.498 & 2.025 \\
\hline
{\bf MD} &--&-0.279& -0.276& -0.279& -0.225& -0.094& -0.160 & -0.163& -0.185& -0.017& -0.013 & -0.399\\
{\bf MAD} &--&0.279& 0.276& 0.279& 0.225& 0.094& 0.160 & 0.163& 0.185& 0.040 &
       0.040 & 0.399 \\
{\bf MRD~\%}& --& -24.8& -24.3& -24.8& -18.3& -11.2& -16.9& -17.3&
       -15.4 &  -2.8&  -2.0& -38.3\\
{\bf MARD~\%} &--&24.8& 24.3& 24.8& 18.3& 11.2& 16.9& 17.3& 15.4 &
        4.7& 4.2& 38.3 \\
\end{tabular}
\end{ruledtabular}

\vspace{1cm}

\begin{ruledtabular}
\begin{tabular}{*{14}{p{1.2cm}}}
    Systems & Ref.& DF &   DF2 &  DF-cx &  DF-cx0 &  DF-cx0-20 &  DF2-B86R  &  DF2-B86R0 &  DF2-B86R0-20 &  DF-optB88 &  DF-optPBE &  DF3-opt1 &  DF3-opt2 \\
\hline
\midrule
1  &0.168&   0.131 &    0.188 & 0.089 & 0.156 &   0.143 &         0.016 &        0.101 &           0.084 &   0.042 &       0.068 &        -0.028 &         0.005 \\
2  &0.343&   0.336 &    0.409 & 0.251 & 0.333 &   0.317 &         0.135 &        0.252 &            0.229 &   0.173 &       0.225 &         0.054 &         0.114 \\
3  & 2.028&   2.003 &    2.088 & 1.951 & 2.143 &   2.106 &         1.850 &        2.064 &           2.022 &   1.893 &       1.918 &         1.787 &         1.833 \\
4  &0.828&   0.930 &    1.032 & 0.822 & 0.926 &   0.906 &         0.686 &        0.832 &           0.803&   0.723 &       0.788 &         0.574 &         0.657 \\
5  &1.650&   1.501 &    1.571 & 1.459 & 1.675 &   1.633 &         1.354 &        1.590 &           1.544 &   1.406 &       1.427 &         1.305 &         1.344 \\
6  &1.594&   1.514 &    1.588 & 1.470 & 1.664 &   1.626 &         1.371 &        1.584 &           1.110 &   1.418 &       1.438 &         1.317 &         1.358 \\
7  &1.570 &   1.446 &    1.517 & 1.397 & 1.599 &   1.561 &         1.283 &        1.508 &          1.464, &   1.344 &       1.366 &         1.232 &         1.270 \\
8  &2.643&   2.425 &    2.523 & 2.392 & 2.640 &   2.592 &         2.282 &        2.552 &          2.500 &   2.345 &       2.351 &         2.232 &         2.268 \\
9  &2.061&   1.973 &    2.057 & 1.919 & 2.121 &   2.082 &         1.803 &        2.031 &            1.987 &   1.860 &       1.885 &         1.744 &         1.787 \\
10 &2.848&   2.679 &    2.782 & 2.644 & 2.886 &   2.839 &         2.547 &        2.808 &            2.757 &   2.599 &       2.605 &         2.495 &         2.531 \\
11 &3.523&   3.301 &    3.382 & 3.269 & 3.609 &   3.542 &         3.183 &        3.549 &          3.477 &   3.237 &       3.237 &         3.147 &         3.172 \\
12 &1.388&   1.391 &    1.476 & 1.323 & 1.500 &   1.465 &         1.201 &        1.404 &            1.364 &   1.261 &       1.289 &         1.132 &         1.181 \\
13 &1.256&   1.209 &    1.297 & 1.143 & 1.324 &   1.289 &         1.006 &        1.217 &         1.175 &   1.079 &       1.108 &         0.941 &         0.988 \\
14 &2.55&   2.361 &    2.471 & 2.310 & 2.560 &   2.511 &         2.173 &        2.462 &          2.406 &   2.253 &       2.268 &         2.110 &         2.154 \\

\hline
{\bf MD}&--& -0.089& -0.005& -0.144&  0.049&  0.012& -0.254& -0.035&  -0.078&   -0.201& -0.177& -0.315& -0.271\\

{\bf MAD}&--& 0.104& 0.073& 0.144& 0.053& 0.037& 0.254& 0.047& 0.078& 0.201& 0.177& 0.315& 0.271 \\

{\bf MRD~\%}& --& -5.0&   3.3& -11.6  &   2.6&  -0.2& -23.0&  -5.9& -9.3& -18.3& -14.7& -29.8& -24.8 \\
{\bf MARD~\%}& --& 6.8&  6.5& 11.6  &  3.9 &  3.7& 23.0&  6.6&  9.3&  18.3& 14.7& 29.8& 24.8\\
\end{tabular}
\end{ruledtabular}
\end{table*}

Table~\ref{tab:table_1} shows the computed data for all studied functionals. The upper part
shows results for all but the \mbox{vdW-DFs}, which is shown in the lower part.

\newpage



\bibliography{ref}

\end{document}